\documentclass[twocolumn,apj,numberedappendix,appendixfloats,twocolappendix]{openjournal}

\usepackage{newtxtext,newtxmath}

\usepackage[T1]{fontenc}

\usepackage{graphicx}
\usepackage[dvipsnames]{xcolor}
\usepackage[colorlinks,linkcolor=RedViolet,citecolor=RedViolet,urlcolor=RedViolet]{hyperref}
\usepackage{amsmath}
\usepackage{soul}
\usepackage{xspace}
\usepackage{ulem}                   
\usepackage{relsize}
\usepackage{orcidlink}

\usepackage{pifont}
\usepackage{orcidlink}
\usepackage{tabularx}

\DeclareRobustCommand{\ion}[2]{%
\relax\ifmmode
\ifx\testbx\f@series
{\mathbf{#1\,\mathsc{#2}}}\else
{\mathrm{#1\,\mathsc{#2}}}\fi
\else\textup{#1\,{\mdseries\textsc{#2}}}%
\fi}

\newcommand{\Msun}{{\rm M_\odot}}

\newcommand{\ie}{{i.e.~}}
\newcommand{\eg}{{e.g.,~}}

\newcommand{\arepo}{{\sc arepo}\xspace}
\newcommand{\areport}{{\sc arepo-rt}\xspace}

\newcommand{\thesan}         {\textsc{thesan}\xspace}
\newcommand{\thesanone}      {\textsc{thesan-1}\xspace}

\newcommand{\thesanzoom}     {\textsc{thesan-zoom}\xspace}
\newcommand{\zf}{4$\times$\xspace}
\newcommand{\ze}{8$\times$\xspace}
\newcommand{\zs}{16$\times$\xspace}
\newcommand{\dd}{\mathrm{d}}

\newcommand{\Mdust}{M_\mathrm{dust}}
\newcommand{\Mstar}{M_\mathrm{star}}
\newcommand{\Tdust}{T_\mathrm{dust}}

\newcommand{\candidates}{all-HR\xspace}

\graphicspath{{./}{figures/}}

\shorttitle{The \thesanzoom project: burstiness and dust survival}
\shortauthors{Garaldi et al.}

\begin{document}
\title{\textbf{The thesan-zoom Project: bursty star formation\\ is incompatible with prolonged dust survival}\vspace{-1.5cm}}
\author{Enrico Garaldi$\orcidlink{0000-0002-6021-7020}^{1,2,3,*}$}
\author{Filip Popovic $\orcidlink{0009-0006-8856-918X}^4$}
\author{Rahul Kannan $\orcidlink{0000-0001-6092-2187}^{4}$}
\author{Aaron Smith $\orcidlink{0000-0002-2838-9033}^{5}$}
\author{Ewald Puchwein $\orcidlink{0000-0001-8778-7587}^{6}$}
\author{Naoki Yoshida $\orcidlink{0000-0001-7925-238X}^{7,1,8}$}
\author{Kentaro Nagamine $\orcidlink{0000-0001-7457-8487}^{9,10,1,11,12}$}
\author{Celine Peroux $\orcidlink{0000-0002-4288-599X}^{13,14}$}
\author{Laura Keating $\orcidlink{0000-0001-5211-1958}^{15}$}
\author{Mark Vogelsberger $\orcidlink{0000-0001-8593-7692}^{16,17}$}
\author{William McClymont $\orcidlink{0009-0009-5565-3790}^{18,19}$}
\author{Xuejian Shen $\orcidlink{0000-0002-6196-823X}^{14}$}
\author{Sandro Tacchella $\orcidlink{0000-0002-8224-4505}^{18,19}$}
\author{Lars Hernquist $\orcidlink{0000-0001-6950-1629}^{20}$\vspace{0.5cm}}
\thanks{$^*$E-mail: \href{mailto:egaraldi@ipmu.jp}{egaraldi@ipmu.jp}}
\affiliation{$^1$ Kavli IPMU (WPI), UTIAS, The University of Tokyo, Kashiwa, Chiba 277-8583, Japan }
\affiliation{$^2$ Center for Data-Driven Discovery, Kavli IPMU (WPI), UTIAS, The University of Tokyo, Kashiwa, Chiba 277-8583, Japan }
\affiliation{$^3$ Institute for Fundamental Physics of the Universe, via Beirut 2, 34151 Trieste, Italy }
\affiliation{$^4$ Department of Physics and Astronomy, York University, 4700 Keele Street, Toronto, ON M3J 1P3, Canada }
\affiliation{$^5$ Department of Physics, The University of Texas at Dallas, Richardson, TX 75080, USA }
\affiliation{$^6$ Leibniz-Institut f\"ur Astrophysik Potsdam, An der Sternwarte 16, 14482 Potsdam, Germany }
\affiliation{$^7$ Department of Physics, School of Science, The University of Tokyo, 7-3-1 Hongo, Bunkyo, Tokyo 113-0033, Japan}
\affiliation{$^8$ Research Center for the Early Universe, School of Science, The University of Tokyo, 7-3-1 Hongo, Bunkyo, Tokyo 113-0033, Japan}
\affiliation{$^9$ Theoretical Astrophysics, Department of Earth and Space Science, The University of Osaka, 1-1 Machikaneyama, Toyonaka, Osaka 560-0043, Japan}
\affiliation{$^{10}$ Theoretical Joint Research, Forefront Research Center, Graduate School of Science, The University of Osaka, Toyonaka, Osaka 560-0043, Japan}
\affiliation{$^{11}$ Department of Physics \& Astronomy, University of Nevada, Las Vegas, 4505 S. Maryland Pkwy, Las Vegas, NV 89154-4002, USA}
\affiliation{$^{12}$ Nevada Center for Astrophysics, University of Nevada, Las Vegas, 4505 S. Maryland Pkwy, Las Vegas, NV 89154-4002, USA}
\affiliation{$^{13}$ European Southern Observatory, Karl-Schwarzschildstrasse 2, D-85748, Garching bei Munchen, Germany}
\affiliation{$^{14}$ Aix Marseille Université, CNRS, LAM (Laboratoire d’Astrophysique de Marseille) UMR 7326, F-13388, Marseille, France}
\affiliation{$^{15}$ Institute for Astronomy, University of Edinburgh, Blackford Hill, Edinburgh, EH9 3HJ, UK }
\affiliation{$^{16}$ Department of Physics, Kavli Institute for Astrophysics and Space Research, Massachusetts Institute of Technology, Cambridge, MA 02139, USA }
\affiliation{$^{17}$Fachbereich Physik, Philipps Universit\"at Marburg, D-35032 Marburg, Germany}
\affiliation{$^{18}$ Kavli Institute for Cosmology, University of Cambridge, Madingley Road, Cambridge CB3 0HA, UK }
\affiliation{$^{19}$Cavendish Laboratory, University of Cambridge, 19 JJ Thomson Avenue, Cambridge CB3 0HE, UK }
\affiliation{$^{20}$ Center for Astrophysics $|$ Harvard $\&$ Smithsonian, 60 Garden Street, Cambridge, MA 02138, USA \vspace{0.2cm}}

\begin{abstract} 
\noindent Cosmic dust is a key regulator of galaxy evolution, but its build-up and survival in the first billion years remain poorly constrained. We present a systematic analysis of dust in the \thesanzoom suite of radiation-hydrodynamical zoom-in simulations, which self-consistently model dust formation, growth, destruction, and its coupling to radiative transfer in galaxies at $z\gtrsim3$, a multi-phase ISM and bursty star formation histories. 

The simulated galaxies reproduce the observed trends of dust-to-gas and dust-to-metal ratios with gas metallicity, while showing a dust deficit at high specific star-formation rates. They also broadly match observed dust temperatures and UV--IR spatial offsets. We find that dust and its properties are strongly time-variable and tightly linked to bursty star formation: during starbursts, young stars heat dust, and stellar feedback rapidly destroys and ejects dust from star-forming regions, while in post-burst phases low gas densities suppress efficient regrowth. This cycle yields short-lived IR-bright phases (median duration of $20.3_{\mathsmaller{-2.4}}^{\mathsmaller{+2.3}}$ Myr) and longer dust-poor phases, naturally producing a correlation between dust temperature and distance from the star-forming main sequence. The predicted attenuation at 1500 \AA{} is low compared to observations ($A_{1500} \lesssim 0.3$ at all stellar masses, increasing to $A_{1500} \lesssim 1$ at $M_\mathrm{star} \sim 10^9 \, \Msun$ when including unresolved dust through post processing), indicating that a mechanism able to shield dust from strong feedback events is necessary to reconcile our galaxy formation model with observations.

In our model, bursty star formation prevents the survival of large dust reservoirs ($M_\mathrm{dust}/M_\mathrm{star} \gtrsim 10^{-3}$) over a significant fraction of cosmic time. This implies that bursty star formation can produce the observed overabundance of UV-bright galaxies at $z \gtrsim 10$ only if it rapidly settles down by $z \sim 8$ (where large dust reservoirs are detected). It is also possible that our models lack physical ingredients or emergent phenomena that aid the survival of dust. Future observations of high-redshift dust will be key to diagnose the physical mechanism at play in the first galaxies.
\end{abstract}
\maketitle

\section{Introduction}
\label{sec:intro}
Cosmic dust is an essential component of many astrophysical systems ranging from galaxies to planetary systems. It consists of small solid particles that reside mainly in interstellar space, having a variety of sizes typically in the range $0.01$ -- $1 \, \mu$m, and dominated in composition by carbon, silicates, and polycyclic aromatic hydrocarbons \citep[\eg][]{Draine2003}. Cosmic dust plays a significant role in the evolution of galaxies by influencing the cooling processes of interstellar gas, as well as by interacting with the radiation field. In particular, the reprocessing of UV radiation into IR wavelengths is key in shaping the epoch of reionization (EoR, \ie the transformation of the intergalactic medium, IGM, from a cold neutral gas into a hot ionized plasma) through its effect on the ionizing photons escape fraction \citep[\eg][]{Yajima+2009, Chisholm+2018, Chisholm+2022, Kostyuk+2023, Pahl+2023}. 
Cosmic dust also acts as a catalyst for molecular hydrogen formation \citep[][]{Hollenbach+McKee1979,Cazaux+Tielens2002,Cazaux+Tielens2004} and provides an additional channel for gas cooling, therefore influencing star formation and gas dynamics. 
\citet{Novak+2017, Fudamoto+2020, Khusanova+2021, Zavala+2021, Algera+2023, vanLeeuwen+2024} find that the majority of the cosmic star formation rate density (SFRD) is dust-obscured at $z \lesssim 4$.

Observations from instruments such as the Atacama Large Millimeter/submillimeter Array (ALMA) and the Northern Extended Millimetre Array (NOEMA) have detected substantial amounts of dust in individual systems up to $z = 8.3$ \citep{Tamura+2019}, indicating rapid dust production, potentially through supernovae or other high-energy stellar events \citep{Watson+2015}. Additionally, samples like ALPINE \citep{alpine}, REBELS \citep{rebels} and ALCS \citep{Jolly+2025} are starting to provide a population-level view of the dust properties of galaxies in the first few Gyr of the Universe \citep[see \eg][]{Sommovigo+2022_ALPINE, Sommovigo+2022_REBELS, Algera+2026}, although still limited to the most massive systems. These findings suggest that, even in the early Universe, dust played an important role in shaping the physical conditions within galaxies. However, the dust continuum has not been detected in galaxies beyond this redshift \citep{Popping&Peroux2022, Bakx+2026}, potentially indicating a drastic change in physical conditions within the interstellar medium (ISM) of galaxies. The dust content of Damped Lyman $\alpha$ (DLA) systems can be estimated up to $z\lesssim5.5$, providing an estimate of the dust mass density in the Universe \citep[\eg][]{deCia+2016, Peroux&Howk2020}.

The advent of the \textit{James Webb Space Telescope} (JWST) has enabled studies of galaxies to unprecedentedly high redshifts. \citet{Markov+2025a} used SED-fitting to derive the dust attenuation law as a function of redshift over the range $2 < z< 12$, showing how the UV bump at 2175 \AA{} observed in the Milky Way emerges over time (see also \citealt{Fisher+2025}, but see \citealt{Witstok+2023_dust}). \citet{Tacchella+2022}, \citet{Maheson+2025} and \citet{Reddy+2026} also performed a similar analysis on a smaller redshift range, but controlling for more galaxy parameters. 
On the other hand, JWST has also identified and characterised massive galaxies with blue UV slopes ($-2 \lesssim \beta \lesssim -2.6$) and small dust attenuations ($A_\mathrm{V} \lesssim 0.02$), indicative of negligible dust content \citep[][sometimes referred to as `blue monsters`]{Castellano+2022, Castellano+2024, Finkelstein+2022, Naidu+2022, Adams+2023, ArrabalHaro+2023, Atek+2023, Bunker+2023, Curtis-Lake+2023, Fujimoto+2023, Fujimoto+2024, Robertson+2023, Robertson+2024, Tacchella+2023, WangB+2023, Carniani+2024, Harikane+2024, Hsiao+2024, Schouws+2025}. These objects are difficult to explain with current galaxy formation models, as such massive galaxies are expected to host a significant amount of dust, which would therefore produce stronger attenuation and redder slopes \citep[see \eg][]{Ferrara+2023}.

Even when dust continuum is detected, inferring the dust properties of galaxies is challenging, as most observations of high-redshift galaxies are carried out in a single band (in the wavelength range relevant for dust physics), creating a strong degeneracy between dust mass, temperature, and emission properties \citep[for an extended discussion see \eg][]{SommovigoAlgera2025}. Constraining the dust temperature in galaxies requires observations of both sides of the dust-induced IR peak \citep[\eg][]{Bakx+2021}, which are rarely available. 

Theoretical models and simulations have become invaluable tools to understand the formation of cosmic structures. The challenges outlined above render them even more important for the study of cosmic dust and its influence on galaxy evolution. 
The majority of large galaxy formation simulations do not self-consistently model cosmic dust. These include, among others, SPHINX \citep{SPHINX,SPHINX20}, FIREbox \citep{SPHINX,SPHINX20}, FLARES \citep{FLARESI}, IllustrisTNG \citep{TNG_Marinacci,TNG_Naiman,TNG_Nelson,TNG_Pillepich,TNG_Springel}, NewHorizon \citep{newhorizon}, FirstLight \citep{FirstLight}, SERRA \citep{serra}, SPICE \citep{SPICE}, CROC \citep{CROC, Esmerian+2022, Esmerian+2024} and FIREbox \citep{firebox, Fireboxhr}. Instead, dust-related works using these simulations rely on post-processing, assuming simple dust compositions and distribution calibrated to some observational quantity. 

In recent years, on-the-fly dust modeling has started to be included in large-volume cosmological simulations, such as THESAN \citep{Thesan_intro, Thesan_igm, Thesan_data, Thesan_Lya},  OBELISK \citep{Obelisk}, CoDaIII \citep{CoDaIII, dustier}, CROCODILE \citep{crocodile}, COLIBRE \citep{colibre, colibre_dust}, Simba-EOR \citep{SimbaEOR}, as well as the simulations presented in \citet{Graziani+2020, diCesare+2023, Hu+2019, Li+2019, Granato+2021}. However, the extremely small scales relevant for dust evolution ($\mu$m-nm for grain-grain interaction and  $\sim0.1$-$1$ pc for the typical size of dense cloud cores) imply that these processes can only be followed through approximate sub-grid models. Even then, the limited resolution of the simulations hinders the accuracy of such prescriptions and often requires additional post-processing corrections or re-calibrations (see \eg \citealt{Parente_dust_review} for a recent review of dust modeling in simulations). Notable attempts at dust modeling include \citet{Bekki2013, Bekki+2015, Li+2021}, who developed some of the first simulations with dust dynamically decoupled from gas (often referred to as `live' dust); \citet{McKinnon+2016, McKinnon+2017, McKinnon+2018}, including an on-the-fly dust evolution model in the moving-mesh code \arepo \citep{Arepo}; \citet{Aoyama+2016, Aoyama+2018, Dubois+2024}, introducing a two-size approximation for the dust grain that enables accounting for shattering and coagulation; \citet{Hirashita&Aoyama2019}, sampling the dust grain size distribution with 30 bins; \citet{Narayanan+2023, Narayanan+2025, Narayanan+2026} combining live dust with a spatially-and temporally-evolving grain size distribution and \citet[][]{Choban+2022, Choban+2025, Choban+2026} who included an on-the-fly dust model in a subset of the FIRE-2 simulations.

In this paper, we present a detailed analysis of the dust properties in the \thesanzoom simulation suite \citep{ThZoom_intro}, a recently-published set of cosmological zoom-in radiation-hydrodynamical simulations designed to study the formation of galaxies during the EoR and until $z=3$. 
They combine high spatial and mass resolution with detailed models of star formation and stellar feedback based on a custom version of the SMUGGLE model \citep{smuggle}, as well as a model for cosmic dust adapted from \citet{McKinnon+2016,McKinnon+2017}, and radiative transfer. \thesanzoom also uses a new technique to accurately include the time- and spatially-varying large-scale radiation field, necessary to capture the coupling between radiation and gas evolution in small galaxies \citep[see \eg][]{ThesanHR, ThZoom_rad}. This allows \thesanzoom to uniquely connect interstellar, circumgalactic, and intergalactic processes, providing a solid realistic theoretical framework for interpreting observations of early galaxies, particularly those emerging from JWST. 

The paper is organized as follows. In Sec.~\ref{sec:methods}, we review the numerical setup. In Sec.~\ref{sec:simulated_properties}, we investigate the dust properties of our simulated galaxies, while in Sec.~\ref{sec:observed_properties}, we perform a simple forward modeling of our data to compare them to observed quantities. Finally, we discuss the implications of our results in the context of the high-redshift Universe in Sec.~\ref{sec:discussion}. We provide concluding remarks in Sec.~\ref{sec:conclusions}.

\section{Methods}
\label{sec:methods}

\subsection{The \thesanzoom simulation suite}
\label{subsec:thesanzoom}

The \thesanzoom simulations are a suite of cosmological radiation-hydrodynamic zoom-in simulations designed to study the formation and evolution of galaxies during the epoch of reionization, first introduced in \citet{ThZoom_intro}. The simulations follow 14 galaxies selected from the large-volume \thesan simulation \citep{Thesan_intro, Thesan_igm, Thesan_data, Thesan_Lya}, logarithmically spaced in halo masses at $z=3$ (the final redshift of the simulation), simulated at three different resolution levels, named \zf, \ze and \zs (note that not all zoom-in regions are simulated at all three resolution levels because of computational limitations, see Table~3 in \citealt{ThZoom_intro}). These correspond (respectively) to a spatial resolution improvement of a factor 4, 8, and 16 with respect to the parent \thesanone box, resulting in a minimum gas softening length of 69.2, 34.6 and 17.3 comoving pc and a gas mass resolution of $9.09 \times 10^3$, $1.14 \times 10^3$ and $1.42 \times 10^2$ $\Msun$.

The simulations employ the \arepo code \citep{Arepo} and include radiative transfer and non-equilibrium primordial thermochemistry according to the \areport module \citep{ArepoRT}, star formation and stellar feedback through a custom version of the SMUGGLE \citep{smuggle} galaxy formation model, and on-the-fly dust physics using an improved version (see Sec.~\ref{subsec:dust_model} for details) of the model developed in \citet{McKinnon+2016, McKinnon+2017}. Note that we do not include black hole modeling in our simulations, since we do not expect Active Galactic Nuclei to impact the properties of galaxies in the mass range covered by our simulations. Crucially, \thesanzoom employs a novel technique to inject the time-varying spatially-inhomogeneous radiation field from the parent \thesanone simulation at the boundary of the high-resolution region of the zoom-in run. This ensures that the impact of inflowing large-scale radiation fields are approximately accounted for, unlike in traditional zoom-in simulations. 

This framework has already enabled a wide range of investigations of early galaxy evolution, and of the connection between the ISM, CGM and IGM. These include studies of star-formation efficiencies in high-redshift galaxies \citep{ThZoom_sfe, ThZoom_sfe_clouds}, of burstiness, quenching and evolution along the star-forming main sequence \citep{ThZoom_bursty, ThZoom_quench}, of long-term imprints of external reionization on galaxy evolution \citep{ThZoom_rad}, of late-time metal-poor star formation \citep{ThZoom_popIII}, of the origin of N/O-enhanced galaxies \citep{ThZoom_NO}, and of the origin of \ion{O}{I} absorbers \citep{ThZoom_OI}. 

In the following, we focus on the dust model and refer the reader to \citet{ThZoom_intro} for details on other aspects of the simulation suite. It is important to note that an emergent property of our galaxy formation model is that it produces bursty star formation histories, with frequent (every few tens to hundreds Myr) and short (few to tens Myr) starbursts, followed by mini-quenched periods of little-to-no star formation, due to the efficient stellar feedback ejecting most of the gas within the galaxy and thus preventing further star formation \citep[for details, see][]{ThZoom_bursty}. Remarkably, this allows us to match the $z \gtrsim 10$ UVLF but results in an excess of galaxies with $-20 \lesssim M_\mathrm{UV} \lesssim -16$ at $z \lesssim 8$ \citep[][Fig.~15]{ThZoom_intro}, although constructing a UVLF from a small number of zoom-in regions is notoriously difficult and prone to large uncertainties.

\subsection{Dust model}
\label{subsec:dust_model}

\begin{figure*}
    \includegraphics[width=\textwidth]{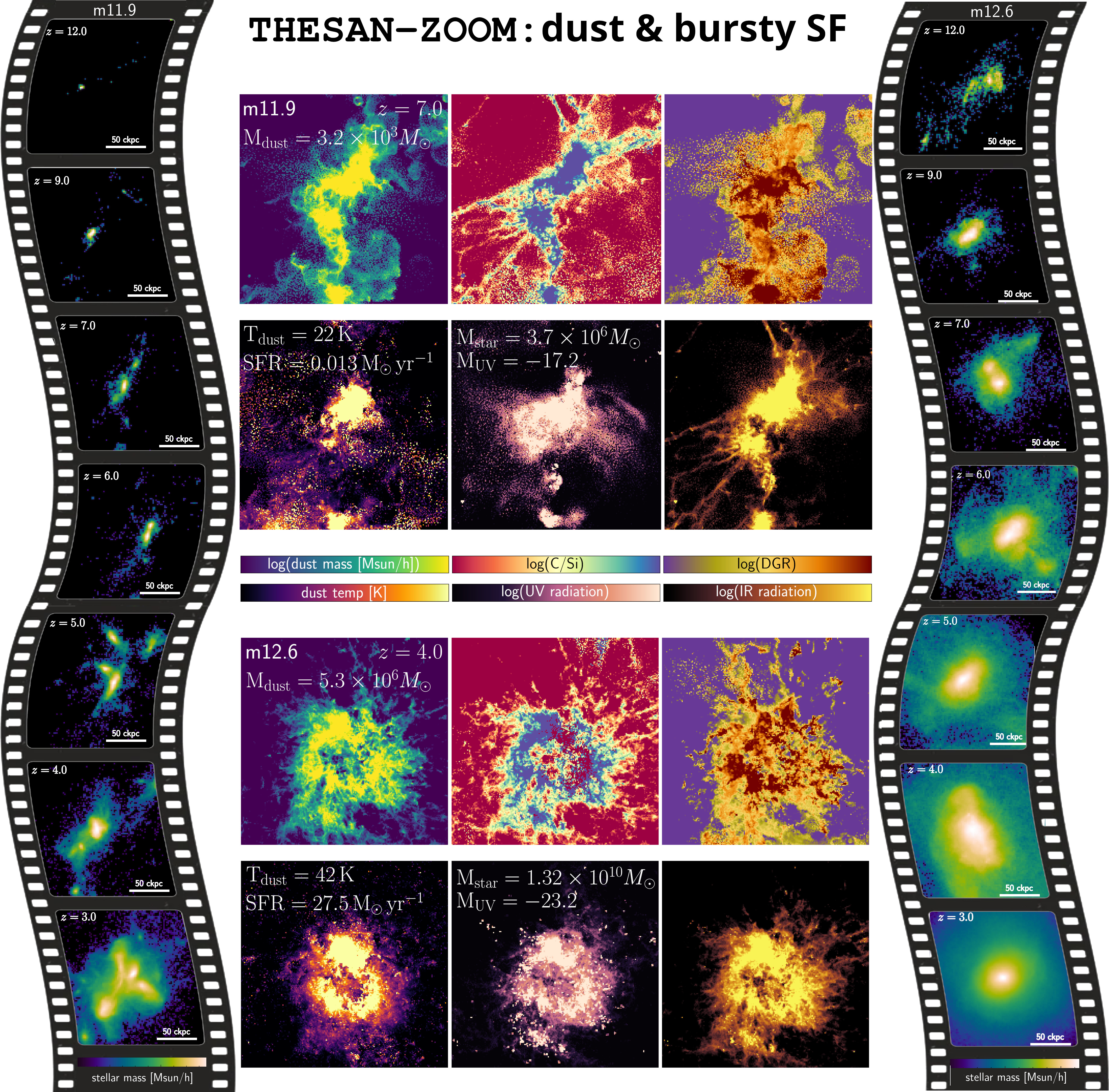}
    \caption{Overview of dust-related properties in two \thesanzoom galaxies. The vertical sequences on the left and right of the plot show the evolution of the projected stellar mass content in the \texttt{m11.9} (left) and \texttt{m12.6} (right) haloes. The size of the panel is set to $140$ comoving kpc/$h$. In the middle part of the Figure, we show two sets of plots (the halo depicted and the redshift are reported in the top left corner). The panels in these plots show (clockwise from the top left) the projected dust mass, dust-to-gas ratio, dust-phase Carbon-to-Silicon ratio, IR and UV radiation fields and dust temperature. All quantities except the dust temperature are log-scaled. The colorscale covers from the 5-th to the 95-th percentile of data. We report in the panels the stellar mass, dust mass, UV magnitude, star formation rate and galaxy-averaged dust temperature.}
    \label{fig:maps}
\end{figure*}

We base our model on the one developed originally in \citet{McKinnon+2016,McKinnon+2017} and calibrated to match local observables. 
We summarize here the salient features of the model and describe the modifications performed to the original version. 

Cosmic dust is treated as a passive scalar, \ie a property of the gas resolution elements. Therefore, this model is by construction unable to capture the drag forces between cosmic dust and gas arising when these two components have a non-zero relative velocity. 
This is a very common approximation \citep[see \eg][]{Bekki2013, Aoyama+2016, dustier, CoDaIII, Graziani+2020, colibre_dust} and it is not expected to have an appreciable impact on our results \citep[see \eg][]{McKinnon+2018, Parente_dust_review}, since the densities resolved by our simulations are not large enough to produce a significant dust-gas drag force. For simplicity and computational efficiency, we further model the dust assuming an effective grain size $a_\mathrm{eff} = 0.1 \mu$m and composition. The effective grain size is chosen to be representative of the dust grains produced by Asymptotic Giant Branch (AGB) stars \citep{Groenewegen1997, Winters+1997, Yasuda+Kozasa2012} and supernovae \citep{Bianchi+Schneider2007, Nozawa+2007}. The main impact of this assumption is on dust accretion, which is grain-size dependent. Appendix B of \citet{McKinnon+2016} demonstrates that the total dust mass is only weakly sensitive to this choice. Given the level of uncertainty in the observed dust properties of galaxies, especially in the redshift range covered by our simulations, we deem this an acceptable compromise. 
Internally, the model tracks independently the dust mass contributed by different elements (namely: Carbon, Oxygen, Magnesium, Silicon, Iron) and production channels (AGB stars, type Ia and typeII supernovae). In order to optimize storage space, in the output we store only the total dust mass and (unlike previous versions of the model) its carbon-to-silicon ratio. The latter allows us to estimate the (local) dust composition by converting the carbon (silicon) mass into a graphite (silicate) abundance. 

In our model, the dust mass is increased through two channels. 
In the first one, AGB stars and supernova explosions inject metals in the surrounding ISM. We assume a fraction of this ejected mass instantly condenses into dust grains. This fraction depends on the carbon-to-oxygen ratio of the gas, as described in \citet{McKinnon+2016}. This is the only mechanism able to create dust in a dust-free environment. 
The second channel is dust growth in the ISM through the deposition of free metals onto dust grains, often simply called `accretion'. In our model, this occurs in star-forming gas\footnote{Note that the definition of star formation in the original \citet{McKinnon+2016, McKinnon+2017} model is different as a consequence of the different resolutions and galaxy formation models. Specifically, in the original \citet{McKinnon+2016, McKinnon+2017} model all gas with density \mbox{$n_\mathrm{gas} \geq 0.106$ cm$^{-3}$} is considered star forming, while in our simulations we employ a Jeans criterion to define such gas \citep[for details see][]{ThZoom_intro}.} with a local timescale that depends on the gas temperature, density, and -- unlike in the original model -- metallicity. Finally, we have newly introduced a gas temperature threshold of $T_\mathrm{gas, thr} = 300$ K for this process, in order to account for the fact that at higher temperatures the large dust-gas relative velocities suppress metal deposition on grain surfaces. In practice, within each cell satisfying the aforementioned temperature condition and at each timestep, we solve for each metal species  the following equation:
\begin{equation}
\begin{array}{rl}
\left( \frac{\dd M_\mathrm{dust, i}}{\dd t} \right)_\mathrm{growth} & =  \left( 1 - \frac{M_\mathrm{dust, i}}{M_\mathrm{metals, i}} \right) \left( \frac{M_\mathrm{dust, i}}{\tau_\mathrm{growth}} \right) \\ 
\tau_\mathrm{growth} & = 3\,\mathrm{Gyr} \, \frac{a_\mathrm{eff}}{0.1 \, \mu \mathrm{m}} \frac{100 \, \mathrm{cm}^{-3}}{\rho_\mathrm{gas}} \left( \frac{20 \mathrm{K}}{T_\mathrm{gas}} \right)^{0.5} \frac{Z_\odot}{Z_\mathrm{gas}}
\end{array}
\end{equation} 
where $\tau_\mathrm{growth}$ is the typical timescale for dust growth, depending on the \textit{local} gas density $\rho_\mathrm{gas}$, temperature $T_\mathrm{gas}$, and metallicity $Z_\mathrm{gas}$. The index $i$ runs over all internally-tracked combinations of elements and production channels. Note that the original \citet{McKinnon+2016} model does not include any metallicity dependence in this growth rate. Following \citet[][]{Hirashita2000} and \citet[][]{Li+2021}, we include a factor $Z_\odot / Z_\mathrm{gas}$ in this timescale. We assume that star-forming gas has a fixed temperature of $10^4$ K \textit{only} for the purpose of computing the dust accretion timescale. We note that this is formally incompatible with the assumption on the maximum ISM temperature allowed, but it allows us to minimize the differences with the original \thesan simulations in star-forming gas that has the same density and metallicity. We have tested that relaxing this assumption has no appreciable effect on our results (see Appendix~\ref{app:newdust}), since in our model dust accretion plays a relatively minor role, as described in Sec.~\ref{sec:discussion}. Thus, future versions of this model will move away from this assumption.

Dust is destroyed through two separate channels. The first is astration, \ie dust within gas cells that collapse into a star, which destroys the dust and transfers its metals to the star. The second channel is thermal and supernova sputtering, \ie the erosion of metals from the surface of dust grains due to collision with energetic particles. In this case, we estimate the dust destruction by solving in each cell the following equation:
\begin{equation}
\left( \frac{\dd M_\mathrm{dust, i}}{\dd t} \right)_\mathrm{sputtering} = - \frac{M_\mathrm{dust, i}}{\tau_\mathrm{sputtering}} 
\end{equation}
where $\tau_\mathrm{sputtering}$ is the typical timescale for sputtering, which is computed differently for thermal and supernova sputtering, namely:
\begin{equation}
\begin{array}{rl}
\tau_\mathrm{sputtering, thermal} & = 0.057 \frac{a_\mathrm{eff}}{0.1 \, \mu \mathrm{m}} \frac{10^{-27}\,\mathrm{cm}^{-3}}{\rho_\mathrm{gas}} \left[ 1 + \left( \frac{2 \times 10^6 \mathrm{K}}{T_\mathrm{gas}}\right)^{2.5} \right] \, \mathrm{Gyr} \\
\tau_\mathrm{sputtering, SN} & = \frac{M_\mathrm{gas}}{\epsilon \gamma M_\mathrm{s,100}}
\end{array}
\end{equation}
where $M_\mathrm{gas}$ is the gas mass in the cell, $\epsilon$ is the grain destruction efficiency by SN shocks, $\gamma$ is the local Type II SN rate, and $M_\mathrm{s,100}$ is the mass of gas shocked to at least 100 \mbox{km s$^{-1}$} \citep[for a derivation of these timescales, see][respectively]{McKinnon+2016, McKinnon+2017}.

Unlike in the original \thesan simulations, dust is fully coupled to gas evolution. Thus, the local dust content contributes to gas cooling and interacts with radiation, reprocessing UV photons into IR ones. Additionally, the coupling to the local radiation field implies that our simulations are able to self-consistently predict the dust temperature $T_\mathrm{dust}$. 
When computing this quantity, we also include the time-evolving contribution from the Cosmic Microwave Background (CMB) radiation, which is not otherwise followed by the radiative transfer scheme. In doing so, we assume that the gas is optically thin to the relevant CMB photons. We note here that this assumption might break down in very high-density ($\gtrsim 10^6$ cm$^{-3}$) metal-rich ($\gtrsim Z_\odot$) and dust-rich regions, allowing the dust to cool below the CMB temperature \citep{Martinez-Gonzalez+2021}. However, we expect such extreme conditions to be rare, especially in a limited sample like ours, and therefore we do not anticipate that this assumption will affect our results. 

We provide an example of simulated galaxies in Fig.~\ref{fig:maps}, where we show as vertical sequences the time evolution of two galaxies (\texttt{m11.9} on the left and \texttt{m12.6} on the right). Each panel in the sequences shows the projected stellar mass in a $140$ comoving kpc/$h$ cube centered on the galaxy. These highlight the variety in formation histories of the \thesanzoom galaxies; \texttt{m11.9} (left) shows a rather eventful history, with multiple large mergers throughout the simulated evolution, while \texttt{m12.6} (right) experiences a much quieter assembly. We select one snapshot for each of these two galaxies and show a set of dust-related properties in the central part of the Figure, again as projections of the same $140$ comoving kpc/$h$ cube. Clockwise from the top left panel, we plot the projected dust mass, dust-to-gas ratio, dust-phase Carbon-to-Silicon ratio (as a proxy of dust composition), the IR and UV radiation fields and the dust temperature. 

Fig.~\ref{fig:maps} shows the complex dust structure in our galaxies, with large spatial variations in dust composition and temperatures, which will be discussed in Sec.~\ref{subsec:temperature}. Another property of the simulated galaxies is the fact that the dust morphology can be quite different than the stellar one, as is particularly evident in the bottom set of panels. We discuss this in Sec.~\ref{sec:offsets}.

\section{Simulated dust properties of the thesan-zoom galaxies}
\label{sec:simulated_properties}
We begin our investigation of the dust properties in \thesanzoom by analysing quantities directly predicted by the simulations, starting with common scaling relations.  
This will serve as a broad assessment of the reliability of the dust content of our simulations. For this reason, here we focus on the fiducial \thesanzoom physical model and defer the investigation of physics variation to a future work.

\subsection{Preliminary considerations}
\label{subsec:preliminary}

For all the figures in the paper, we employ the same color-coding, namely:
\begin{itemize}
    \item the colored symbols represent the target galaxies of \thesanzoom simulations (referred to as `targets'), 
    \item the background grey histogram shows the distribution obtained from non-target galaxies completely within the high-resolution region of the simulation and free from contamination from low-resolution particles (referred to as `\candidates');
    \item the purple contour indicates the central 68\% of the data in the original \thesan simulation.
\end{itemize}

In each panel we include targets and \candidates from all snapshots within the redshift range covered; thus, each panel contains multiple data points with the same color and shape. 
The number of \candidates typically increases until $z\sim7$ and then decreases towards lower redshifts. This is due to the combination of two effects. First, the process of structure formation produces more and more objects with time. Simultaneously, our high-resolution region is defined from the Lagrangian region of the target, so the volume available for non-target haloes shrinks with time. In other words, the high-resolution region is, by construction, collapsing (almost) entirely onto the target halo by $z=3$. 

Most observed dust properties are inferred from infrared observations that are dominated by the hot and bright dust. 
This is a complex problem that requires, in principle, full forward-modeling through infrared radiative transfer to enable a faithful comparison with observations. 
However, given the exploratory nature of this paper, we rely on a simpler approach and approximately compensate for this observational bias by computing intensive dust properties of simulated galaxies using only star-forming gas cells, where the dust temperature is expected to be larger and, therefore, dominate the spectral energy distribution of the galaxy. Extensive properties are instead computed using all gas cells within twice the stellar half-mass radius. In a forthcoming paper (Popovic et al. in prep) we will provide \texttt{SKIRT}-based \citep{skirt} modeling of the simulated galaxies.

\subsubsection{Observations used in the paper}
\label{subsec:observation_descr}
In this paper we will compare the simulated data with a wide range of observations. For the sake of clarity, we provide here an overview of the main observational techniques used in the relevant works.

\emph{Panchromatic SED modeling.} 
A common technique to characterize the integrated dust properties of star-forming galaxies is `panchromatic' SED modeling across the UV, optical, and IR wavelength \citep{Remy-Ruyer+2013, Rowlands+2014, daCunha2015, Burgarella+2020, Burgarella+2025, Burgarella+2026}. This often includes constraints from energy balance, enforcing that radiation energy absorbed by dust at UV and optical wavelengths must be re-emitted as thermal radiation in the far-IR. This modeling allows for the simultaneous determination of stellar mass, star formation rates, and dust attenuation, while loosely constraining dust mass and luminosity-weighted temperatures from the IR emission peak. Recently, studies have began to combine photometry with spectroscopic line ratios to improve estimates of dust attenuation and the resulting IR luminosity. While powerful (as they can be applied to photometric observations with limited coverage), this technique suffers from a number of limitations. There is a degeneracy between the inferred stellar age, dust attenuation and photometric redshift, particularly in highly-obscured regimes. The energy balance assumption breaks down when the peaks of UV and IR radiations are physically segregated (as it is often the case, see discussion in Sec.~\ref{sec:offsets}). Finally, results can be biased by the presence of an AGN or if the templates used in the SED fitting procedure do not capture the true physical conditions. 

\emph{Interferometric and single-dish submillimeter observations.} 
A number of studies focus on the direct detection of thermal dust emission through observations at millimeter  and sub-mm wavelengths. Facilities such as ALMA and NOEMA have revolutionized this field by providing the sensitivity to detect dust in `normal' star-forming galaxies at $z>6$ \citep{Mancini+2015, Schaerer+2014,Watson+2015, Witstok+2023, Algera+2024}. These observations typically probe the Rayleigh-Jeans tail of the dust SED, where the emission is generally considered to be optically thin. Under this assumption, the measured continuum flux serves as a direct tracer of the total dust mass, provided a mass-weighted dust temperature is known (or assumed, as often done). The dust temperature-mass degeneracy is one of the main limiting factors for the accuracy of these studies. A limited number of studies has focused on multi-band observations, in order to better constrain the shape of the IR dust emission peak, and therefore to break the degeneracy between dust mass and temperature. Other limitations include: instrument sensitivity limiting observations to the brightest sources; the presence of optically-thick regions mimicking the effect of larger dust masses \citep[\eg][]{Algera+2024}; and uncertainties in the dust absorption coefficients and emissivity indices, producing uncertainties of up to 0.7 dex in derived masses \citep[\eg][]{Mancini+2015}.

\emph{Absorption-line spectroscopy and metal depletion.} 
A complementary technique to study the ISM of galaxies is absorption-line spectroscopy using quasars as backlights \citep{deCia+2016, Wiseman+2017}. By identifying galaxies as Damped Lyman-$\alpha$ absorbers (DLAs) and measuring the column densities of various ions in the gas phase, one can infer the metal depletion due to refractory elements being locked into solid dust grains. Elements with high condensation temperatures (\eg Fe, Cr, Ni) show significantly lower abundances compared to non-refractory elements (\eg Zn, P, S). The relative abundance ratio [Zn/Fe] is frequently utilized as a reliable tracer of the overall dust content and the depletion strength in the ISM. While the aforementioned methods provide galaxy-averaged quantities, this techniques probes individual lines of sight through the ISM. 
A primary challenge for this technique is disentangling gas-phase abundance variations caused by the galaxy star formation history (through nucleosynthesis) from those caused by metals deposition onto dust grains. Moreover, critical elements for dust estimation are often difficult to constrain because their primary UV absorption lines are frequently saturated \citep{deCia+2016}.

\emph{Stacking, lensing, and multi-line constraints.} 
Source stacking and gravitational lensing are often used to overcome the sensitivity limits of current observations. It should be noted that both techniques also have disadvantages. In the case of stacking, the results are biased towards the brightest sources in the sample (which, for dust observations, translates to those with larger dust temperature) and are prone to contamination from interlopers (\ie galaxies with mis-identified redshift). In the case of lensing, uncertainties in the lensing potential reconstruction propagate to the inferred physical properties. This was done in, \eg \citet{Jolly+2025}.

Novel methods have been developed to break the degeneracy between dust temperature and mass that often plagues single-band far-IR observations, \eg combining continuum flux observations with overlapping fine-structure lines, such as [C II] 158 $\mu$m or [O III] 88 $\mu$m. This enables to better infer dust temperatures and thus reduce the degeneracy with dust masses, but often rely on conversion factors and dust-to-gas ratio scalings calibrated from the local Universe or from numerical simulations, which may not hold for high-redshift galaxies. Such approach was used by, \eg \citet{Sommovigo+2022_ALPINE, Sommovigo+2022_REBELS}.

\subsection{Dust masses}
\label{subsec:dust_mass}

\begin{figure*}
    \includegraphics[width=0.95\textwidth]{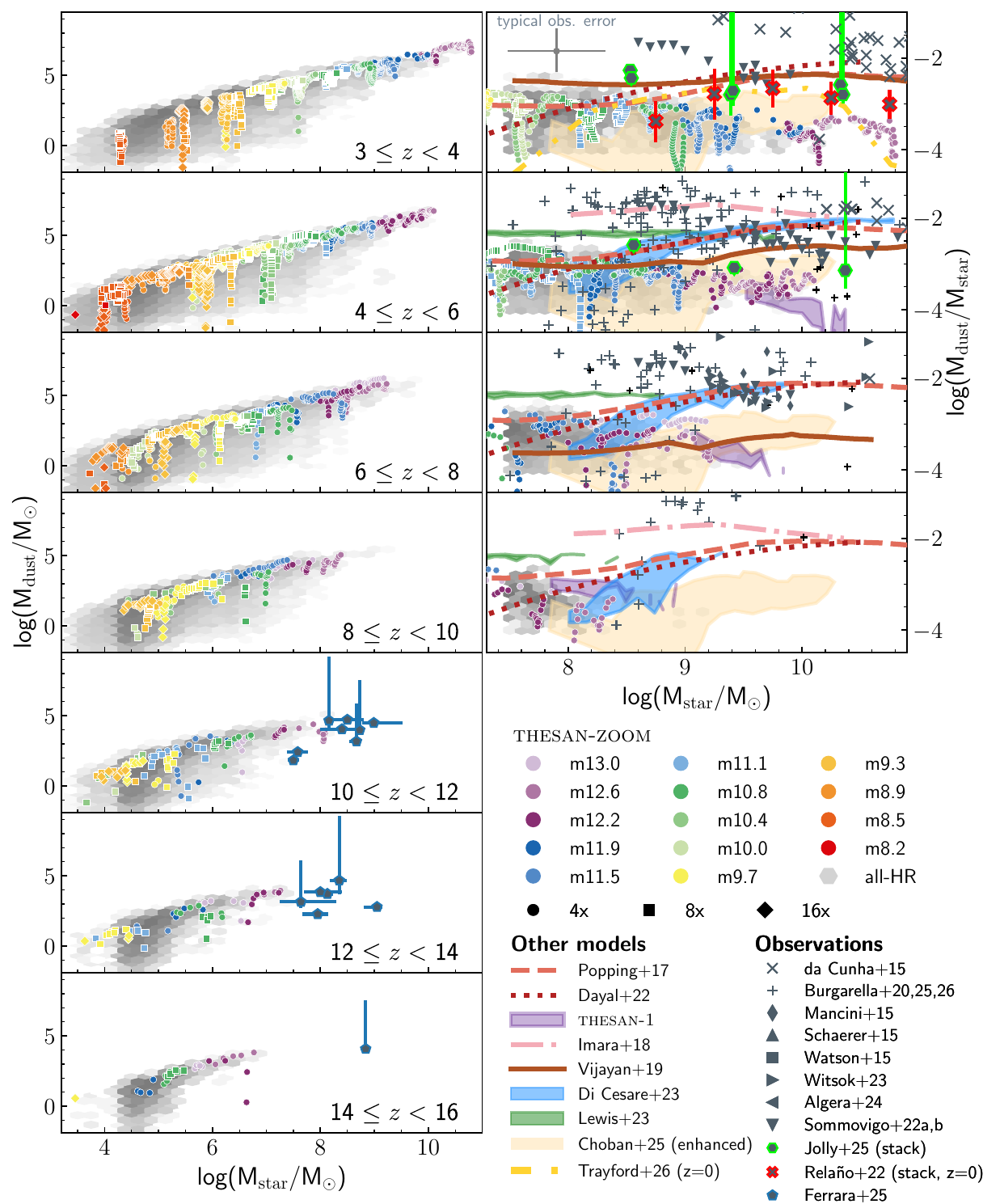}
    \caption{\textbf{Dust-to-stellar mass relation for \thesanzoom galaxies}. Each row shows a different redshift range. Panels on the right side show zoom-ins of the high M$_\mathrm{star}$ end, with the vertical axis normalized by M$_\mathrm{star}$ for visual clarity. 
    Different colored symbols represent different targets and resolution levels. The gray background histogram shows simulated \candidates. Dark grey symbols report individual observations from \citet[][crosses]{daCunha2015}, \citet[][plus symbol]{Burgarella+2020, Burgarella+2025, Burgarella+2026}  , \citet[][diamonds]{Mancini+2015}, \citet[][upward-pointing triangles]{Schaerer+2014}, \citet[][squares]{Watson+2015}, \citet[][right-pointing triangles]{Witstok+2023}, \citet[][left-pointing triangles]{Algera+2024}, \citet[][downward-pointing triangles]{Sommovigo+2022_ALPINE, Sommovigo+2022_REBELS}, as well as the collection of `blue monsters' from \citet[][pentagons with blue outline; notice that the dust mass is derived from their UV attenuation]{Ferrara+2025_bluemonster}. We also show population-averaged results from \citet[][for $z=0$, `X's with red outline]{Relano+2022} and from the stacking analysis of \citet[][hexagons with green outline]{Jolly+2025}. We omit errorbars for visual clarity, but report the typical observational error in the top right panel for reference. The thick orange dashed, red dotted, pink dot-dashed and brown solid lines show the predictions of the models of \citet{Popping+17}, \citet{Dayal+2022}, \citet{Imara+2018} and \citet{Vijayan+2019}, respectively. The purple, green, blue and yellow areas report results from, respectively, the original \thesan simulation (central 68\% of the data), the simulations in \citet[][central 68\% of the data]{diCesare+2023}, \citet[][central 50\% of the data]{dustier}, \citet[][`enhanced' contours in their Fig.~6]{Choban+2025}. The violet double-dot-dashed line shows the results from \citet[][]{colibre_dust}. \textbf{\thesanzoom galaxies underpredict the dust content of massive galaxies, but agree well with the dust-poor population found by \citet{Burgarella+2025, Burgarella+2026} and the stacking analysis of \citet{Jolly+2025}.} }
    \label{fig:Mdust_vs_Mstar}
\end{figure*}

The first relation that we explore, shown in Fig.~\ref{fig:Mdust_vs_Mstar}, is the scaling of dust mass in a galaxy with its stellar content. In this case, both quantities are measured within the stellar half-mass radius for consistency. Most \thesanzoom galaxies lie on an approximate power-law scaling with $\Mdust \approx 0.001 \Mstar$. However, there is a large scatter around such relation, especially in low mass systems. This is driven by the bursty star formation in our model \citep[\eg][]{ThZoom_bursty}, efficiently removing dust (\ie destroying or ejecting it) through stellar feedback. This can be seen in the vertical stripes of same-color symbols tracing the rapid dust growth at fixed stellar mass, and will be further discussed below. The effect is reduced towards larger systems where the deeper gravitational potential can more effectively preserve dust and gas bound to the system or quickly form new dust after it has been destroyed. Notice that from this analysis it is not possible to disentangle dust destruction from simple removal from the halo. We will show later that, in our model, dust is mostly destroyed by the stellar feedback.

At large stellar masses, our simulations underestimate the dust content of galaxies observed by \citet[][plus symbol]{Burgarella+2020, Burgarella+2025, Burgarella+2026}, \citet[][diamonds]{Mancini+2015}, \citet[][triangles]{Schaerer+2014}, \citet[][squares]{Watson+2015}, \citet[][left-pointing triangles]{Algera+2026}, \citet[][crosses]{daCunha2015}, \citet[][right-pointing triangles]{Witstok+2023}, and \citet[][downward-pointing triangles]{Sommovigo+2022_ALPINE, Sommovigo+2022_REBELS}. However, the results from the stacking analysis of \citet[][hexagons with green outline]{Jolly+2025} and \citet[][`X's with red outline, at $z=0$]{Relano+2022} are much closer to our predictions, suggesting that observations might be biased towards dust-rich galaxies, as they are easier to detect in dust continuum. This could explain, at least in part, the difference between our model and observations. 

It is interesting to focus on the second row ($z\sim5$) of Fig.~\ref{fig:Mdust_vs_Mstar}, where we find that our galaxies have approximately one order of magnitude less dust than the one reported in \citet{daCunha2015} and \citet[][downward-pointing triangles]{Sommovigo+2022_ALPINE}, but are consistent with the data from \citet{Burgarella+2025, Burgarella+2026}. These authors identify two populations of galaxies, one hosting large amounts of dust and another harboring much less of this component. This can be seen in the panel as two branches that split at approximately $\log(M_\mathrm{star} / M_\odot) \lesssim 9.5$. Our \thesanzoom galaxies are compatible with the lower branch, but fall short of the upper one. 
Remarkably, our simulations seem to cover a similar parameter space as the `blue monster' of \citet[][pentagon with blue outline; notice that the dust mass is derived from their UV attenuation]{Ferrara+2025_bluemonster}. We will discuss in detail the comparison with both these datasets in Sec.~\ref{sec:discussion}, after presenting a comprehensive view of our simulations. 

We also compare the \thesanzoom predictions with other theoretical works. 
First, comparing our results with the original \thesan simulation (purple contours in the Figure) reveals a good match at $z \gtrsim 9$, while at lower redshifts the \thesanzoom galaxies are more dusty for a given stellar mass. This is due to the much higher resolution of \thesanzoom (between 64 and 4000 times better in mass resolution, depending on the zoom level), allowing us to resolve denser gas, where dust grain accretion is more efficient. We find that our result agrees well with the ones from the FIRE2-based simulations in \citet[][we report with a brown contour the region corresponding to their `enhanced' model in their Fig.~6]{Choban+2025}. Such a flat evolution is predicted also by the \texttt{dustier} \citep{dustier} simulations, although with a higher normalization, and by the semi-analytical model of \citet[][pink dot-dashed line]{Imara+2018}. 
Finally, the semi-analytical models of \citet[][using their fiducial model, orange dashed line in the Figure]{Popping+17} and \citet[][red dotted line]{Dayal+2022}, the \texttt{dustyGadget} simulations in \citet{Graziani+2020, diCesare+2023}, as well as the COLIBRE simulations \citet[][violet double-dot-dashed line showing results at $z=0$]{colibre_dust} present an increasing dust-to-stellar-mass ratio at all redshifts, so that their predictions agree with our simulations at the low-mass end but progressively diverge for more massive objects. The semi-analytical model of \citet[][solid brown line]{Vijayan+2019}, instead, predicts a flat trend at $z\sim7$ and a steeper one at $z\sim5$. 

The results presented above show that \thesanzoom objects represent a relatively dust-poor galaxy population \citep[akin to the one reported in ][]{Burgarella+2025, Burgarella+2026}.  
A possible reason for this is the explosive feedback that characterizes our model \citep{ThZoom_bursty}, which promotes efficient dust destruction by supernovae, hindering the build-up of dust in the galaxies. 
Alternatively, it is possible that our resolution and/or galaxy formation model does not allow for the survival of enough cold gas clumps, where dust can efficiently accrete metals from the ISM. 
In the first case, we expect the dust-to-stellar mass ratio to underestimate observations only for galaxies with a large specific star formation rate (sSFR). If, instead, the lack of cold gas clumps is to be blamed, then we expect dust-to-stellar mass ratio to be underestimated at all sSFR. We do not find direct evidence of the latter in our resolution tests (see Appendix~\ref{app:numerical_convergence}, but it is possible that even our highest-resolution runs do not reach a sufficiently-high resolution. 
For completeness, we mention that it is possible that the timescale for dust growth in the ISM or the stellar yields are underestimated. The latter are unchanged with respect to the original model of \citet{McKinnon+2016, McKinnon+2017}, which is able to match Milky Way data, while the former has been used in multiple works \citep[\eg][]{Li+2021} that show a good agreement with observations. Therefore, we consider this explanation to be less likely than the other two. 

\begin{figure*}
    \includegraphics[width=0.9\textwidth]{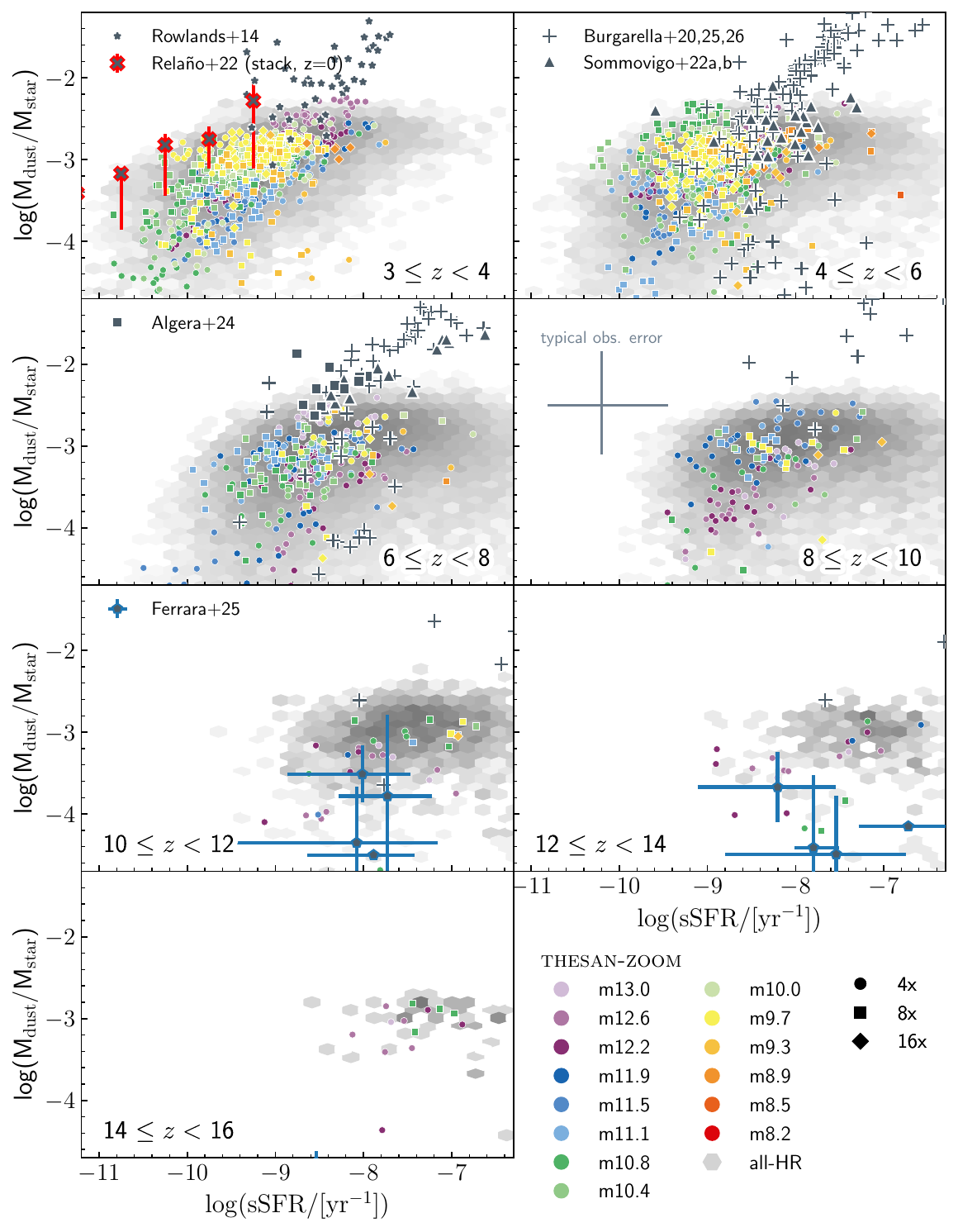}
    \caption{\textbf{Dust-to-stellar mass ratio vs. specific star-formation rate (sSFR) for \thesanzoom galaxies}. Each panel shows a different redshift range. Different colors represent different targets while symbols reflect their resolution level. The gray background histogram shows non-target simulated galaxies. 
    Symbols show observations from \citet{Rowlands+2014}, \citet{Burgarella+2020,Burgarella+2025, Burgarella+2026}, \citet{Sommovigo+2022_REBELS,Sommovigo+2022_ALPINE}, \citet[][]{Algera+2024}, the population-averaged results at $z=0$ from \citet{Relano+2022}, as well as the collection of `blue monsters' from \citet[][notice that the dust mass is derived from their UV attenuation]{Ferrara+2025_bluemonster}. The \thesanzoom galaxies match observations at all but the highest sSFR, including the `blue monsters'. \textbf{At high sSFR dust is efficiently removed from our simulated galaxies, failing short of observational data.}}
    \label{fig:dust-to-star_vs_SFR}
\end{figure*}

In order to investigate the reason behind the low dust content in our simulated massive galaxies, we plot in Fig.~\ref{fig:dust-to-star_vs_SFR} the dust-to-stellar mass ratio as a function of the specific star-formation rate (sSFR) of galaxies. We also show observations from \citet[][stars]{Rowlands+2014}, \citet[][pentagons]{Burgarella+2020}, \citet[][plus symbols]{Burgarella+2025, Burgarella+2026}, \citet[][triangles]{Sommovigo+2022_REBELS}, \citet[][diamonds]{Sommovigo+2022_ALPINE}, \citet[][squares]{Algera+2026}, the population-averaged results at $z=0$ from \citet{Relano+2022} and the `blue monsters' collected by \citet[][notice that the dust mass is derived from their UV attenuation]{Ferrara+2025_bluemonster}. 

It is clear from the Figure (and especially in the panels covering $4 \leq z < 6$ and $6 \leq z < 8$) that our \thesanzoom galaxies reproduce very well the $\Mdust/\Mstar$ values observed at $\log(\mathrm{sSFR}/\mathrm{yr}^{-1}) \lesssim -8.5$, but progressively diverge at higher sSFR, predicting dust-to-stellar mass ratios that are significantly below the observed ones. Thus, we consider this evidence that the bursty nature of the \thesanzoom galaxy formation model is the culprit for the low dust content of massive galaxies.

\subsection{Dust mass function}
\label{subsec:dmf}

\begin{figure}
    \includegraphics[width=\columnwidth]{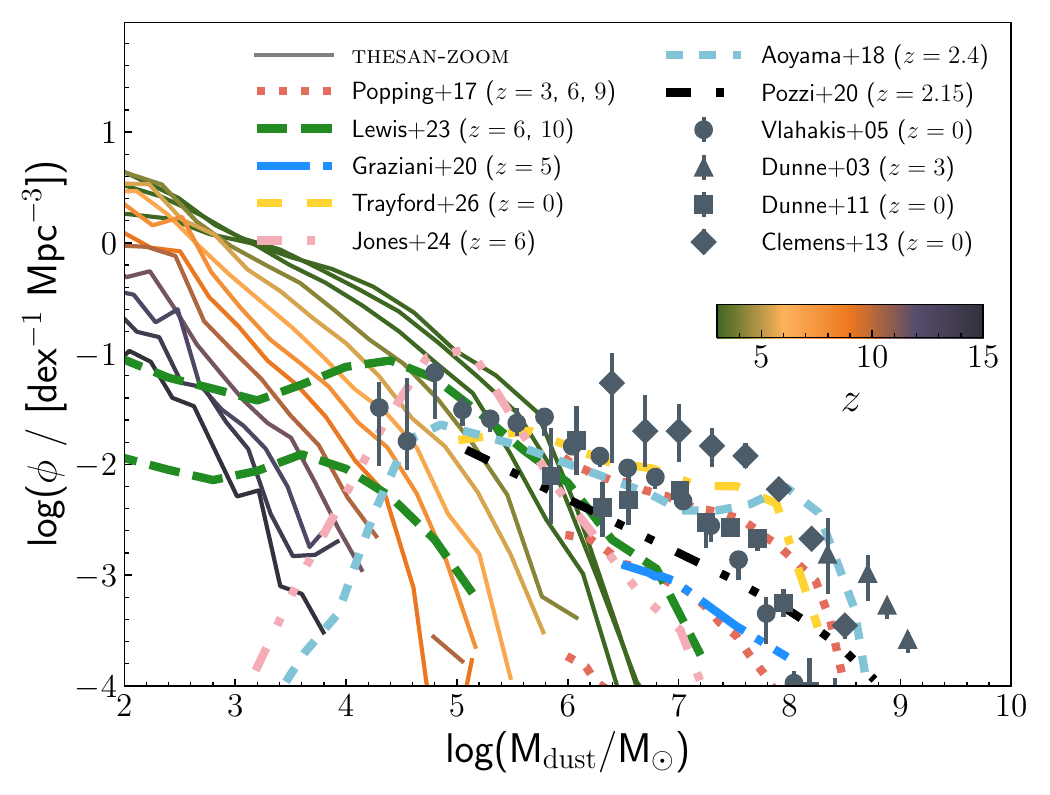}
    \caption{\textbf{Dust mass function} estimated from \thesanzoom (see text for details on how it is computed) over the redshift range $3 \le z \le 15$ (colored solid lines), compared to models from \citet[][dotted orange line]{Popping+17}, \citet[][long-dashed green line]{dustier}, \citet[][dot-dashed blue line]{Graziani+2020}, \citet[][double-dashed-dotted yellow line]{colibre_dust}, \citet[][triple-dot-dashed pink line]{SimbaEOR} and \citet[][short-dashed light blue line]{Aoyama+2018}, as well as observations from \citet[][double-dot-dashed black line]{Pozzi+2020}, \citet[][grey circles]{Vlahakis+2005}, \citet[][grey triangles]{Dunne+2003}, \citet[][grey squares]{Dunne+2011} and \citet[][grey diamonds]{Clemens+2013}.}
    \label{fig:DMF}
\end{figure}

In order to ease a comparison with observations, where stellar masses are often uncertain at the highest redshifts, we build an approximate dust mass function (DMF). Since \thesanzoom includes only a limited number of objects, we follow the same approach used in \citet{ThZoom_intro}, which is adapted from \citet{Ma+2018}. 
In practice, at each redshift, we separate all high-resolution 
haloes in mass bins, each containing $N_\mathrm{sim} (M_\mathrm{halo})$ simulated galaxies. This number depends entirely on our selection of initial conditions. Then, we estimate the expected number of haloes $N_\mathrm{hmf} (M_\mathrm{halo})$ in each mass bin using the halo mass function of \citet{Behroozi+2013}, which agrees very well with results from the parent simulation of \thesanzoom but avoids sampling variance at the massive end. The ratio $N_\mathrm{sim}/N_\mathrm{hmf}$ provides a mass-dependent correction factor that compensates for the selection function of simulated zoom-in galaxies. Thus, any global statistical quantity (\eg mass and luminosity functions) can be computed as usual from the simulated sample by including this correction factor. 
As discussed in \citet{ThZoom_intro}, this method returns the expected stellar mass function as long as the halo selection is not strongly biased. \thesanzoom target galaxies are selected simply based on their masses and therefore should not suffer from such issues. Other \candidates galaxies, however, are by construction close to a larger object and therefore could be somewhat biased if their star formation efficiency is affected by such proximity. 

The recovered dust mass function is shown in Fig.~\ref{fig:DMF} at different redshifts using solid lines. As expected, the DMF builds up steadily over time. We compare our results with a number of other numerical \citep{dustier, Graziani+2020, Pozzi+2020} and semi-numerical \citep{Popping+17} works. In general, we predict a comparable number density of galaxies in the range $\log(M_\mathrm{dust} / \Msun) \sim 4.5$ -- $6.5$, but predict a much smaller number of galaxies with larger dust masses. This effect cannot be fully accounted for by the limited volume of our simulations, since the results from \citet{dustier} and \citet{Graziani+2020} are based on comparable volumes at comparable redshifts. Instead, this discrepancy just reflects the low dust-to-stellar ratio in \thesanzoom galaxies discussed in the previous Section. 

For completeness, we also show in Fig.~\ref{fig:DMF} observations from  \citet{Vlahakis+2005}, \citet{Dunne+2003, Dunne+2011} and \citet{Clemens+2013}. However, reliable observations of the DMF are only available at $z\approx0$ or for galaxies orders of magnitude more massive than those covered in \thesanzoom, preventing a meaningful comparison.

\subsection{Dust-to-gas ratio}
\label{subsec:dust-to-gas}

\begin{figure*}
    \includegraphics[width=0.9\textwidth]{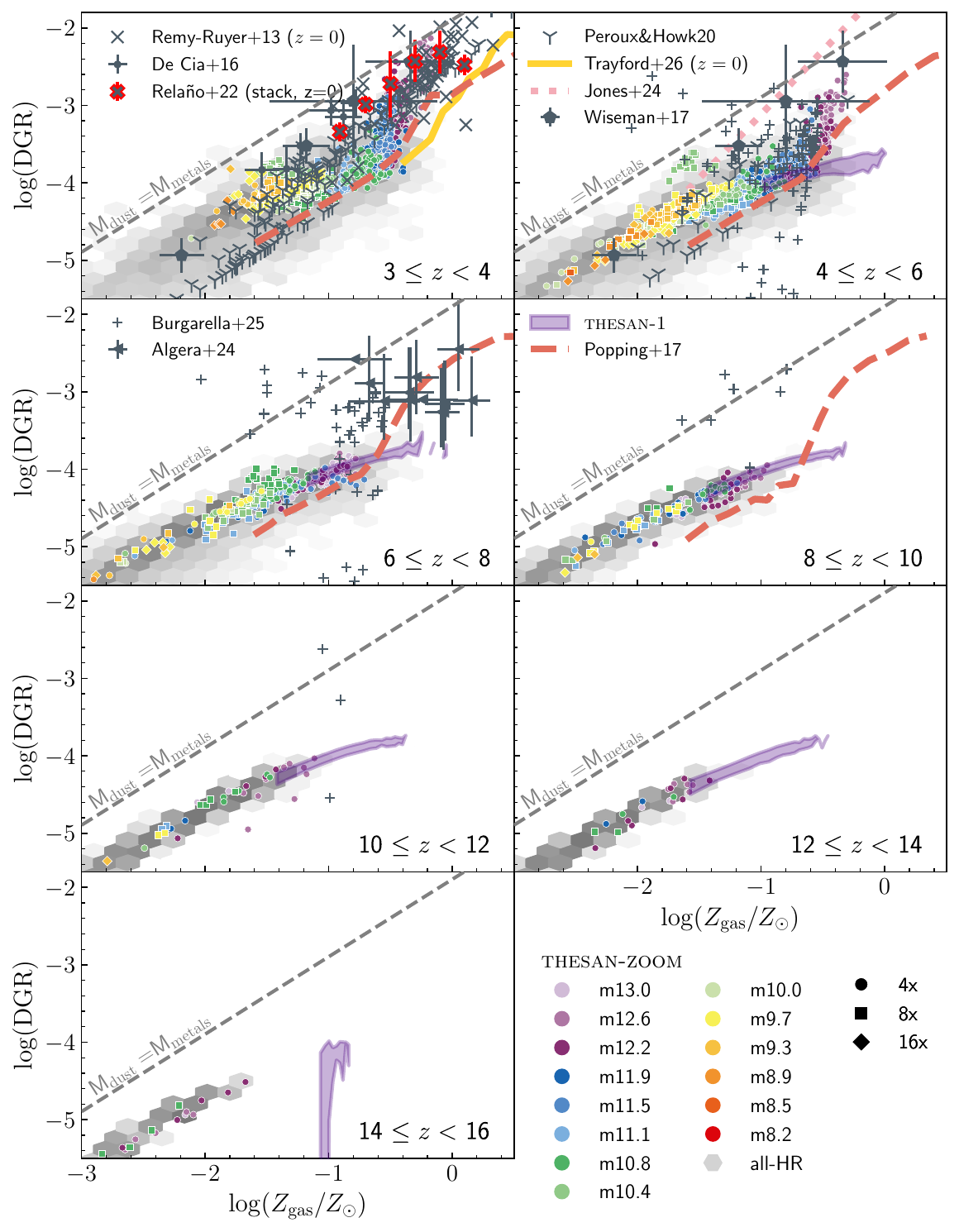}
    \caption{\textbf{Dust-to-gas mass ratio (DGR) as a function of gas-phase metallicity for \thesanzoom galaxies}. Each panel shows a different redshift range. Different colors represent different targets while symbols reflect their resolution level. The gray background histogram shows non-target simulated galaxies. The thick purple contour shows the central 68\% of the data from the original \thesan simulation. Black symbols show observations from \citet[][crosses]{Remy-Ruyer+2013}, 
    \citet[][pentagons]{Wiseman+2017}, \citet[][circles]{deCia+2016}, \citet[plus symbols][]{Burgarella+2025}, \citet[][tripods]{Peroux&Howk2020}, and the population-averaged results at $z=0$ from \citet{Relano+2022}. The orange dashed line shows predictions from the semi-analytical model of \citet{Popping+17}, while the purple solid and red dotted lines show the results from the hydrodynamical simulations in \citet[][COLIBRE]{colibre_dust} and \citet[][Simba-EoR]{SimbaEOR}, respectively. The grey dashed line shows the theoretical maximum DGR achievable, corresponding to all metals in a galaxy locked in dust grains. \textbf{The simulated galaxies agree well with most observations}, with the exception of the recent data from \citet{Burgarella+2025}.}
    \label{fig:DGR_vs_Zgas}
\end{figure*}

A common scaling relation employed to assess the dust content of galaxies and its relation with other baryonic components is the relation between the dust-to-gas (mass) ratio (DGR) and the gas-phase metallicity ($Z_\mathrm{gas}$). This is useful to reveal the relative importance of different dust growth regimes \citep[\eg][]{Popping+17}. 
We show this relation in Fig.~\ref{fig:DGR_vs_Zgas}. 
There are two evident features. First, our model predicts a tight relation between the quantities in the two axes at all redshifts. Second, this relation changes rather abruptly at $\log(Z_\mathrm{gas}) \gtrsim -0.5$. Both features have a common explanation, namely that at lower metallicities, dust growth is dominated by stellar yields in our model, but dust accretion in the ISM becomes more and more efficient (and quick) as the metallicity increases, steepening the relation \citep[\eg][]{Popping+17}. Eventually, once most of the gas-phase metals are locked into dust, this accretion channel is inhibited and the relation is restored with a different normalization \citep[see \eg][]{Aoyama+2018}. We do not see such behaviour in our simulations, except potentially in the most metal-rich objects at $3 \leq z < 4$. This is (at least partially) due to the fact that it takes approximately 1 Gyr to reach such saturation regime \citep{Aoyama+2020}.

Our simulations show an overall agreement with available observations from \citet[][crosses, at $z=0$]{Remy-Ruyer+2013}, 
\citet[][pentagons]{Wiseman+2017}, \citet[][circles]{deCia+2016}, \citet[][tripods]{Peroux&Howk2020}, and \citet[][crosses with red outline, showing population-averaged values]{Relano+2022}, especially considering the discrepancies discussed in Sec.~\ref{subsec:dust_mass}. 
Nevertheless, there are criticalities. First, our simulated values are systematically on the lower end of the observed ones. Second, the DGR values inferred by \citet{deCia+2016} and \citet{Wiseman+2017} do not exhibit any feature at $Z_\mathrm{gas} \approx 0.1 Z_\odot$, where instead both our simulations and the semi-analytical model of \citet{Popping+17} show a change in slope. 
It is important to realize that these authors inferred the DGR values from absorption features in quasar or Gamma Ray Burst spectra, therefore probing the dust column density of individual lines of sight instead of galaxy-averaged quantities as in the case of other reported values. It is therefore possible that such a radically different observational approach leaves an impact on the inferred values. We investigate this in detail in Sec.~\ref{sec:dsddf} through appropriate forward modeling. However the population-averaged values from \citet{Relano+2022}, although based on $z\approx0$ galaxies, also lack any feature at $Z_\mathrm{gas} \approx 0.1 Z_\odot$, thus disfavouring selection effects.

The comparison with the data from \citet{Burgarella+2025} is more puzzling. The two galaxy populations that they identified are located approximately a factor of 10 above and below the simulated relation. Thus, either the observations in \citet{Burgarella+2025} are biased, potentially as a consequence of using a photometric sample with limited band coverage, or they describe galaxies that are fundamentally different from those produced in \thesanzoom.

It is interesting to note that our simulated DGR is in much better agreement with data than the dust-to-stellar ratio (see Fig.~\ref{fig:Mdust_vs_Mstar} and relative discussion). This implies that the \thesanzoom galaxies are also gas-poor. In fact, as we thoroughly describe in Sec.~\ref{subsec:burstiness_and_dust_survival}, our galaxy formation model ejects large amounts of dust and gas through stellar feedback, creating dust- and gas-poor galaxies.

Finally, we show in Fig.~\ref{fig:DGR_vs_Zgas} the results from the original \thesan simulation (purple contour showing the central 68\% of the data, only available at $z \geq 5.5$). Our simulations agree well with these data below the critical metallicity for ISM growth to become efficient, but diverge afterwards, with \thesanzoom providing much better agreement with data. The reason for such a difference is twofold, namely: the new metallicity dependence in the ISM accretion timescale (see Sec.~\ref{sec:methods}) and the much higher resolution achieved, enabling the modeling of cold gas clumps where accretion is efficient \citep[see \eg][]{ThZoom_intro}. 

We also show results from two other large-volume simulations, namely COLIBRE \citep[][solid purple line, showing their results at $z=0$]{colibre_dust} and Simba-EoR \citep[][dotted red line, $z=5.5$]{SimbaEOR}. The former predict dust-poorer galaxies than ours, although the comparison is difficult because of the large differences in redshift covered. Simba-EoR, instead, predicts galaxies that are dustier at most metallicites than in our \thesanzoom model. 

\subsection{Dust-to-metal ratio}
\label{subsec:dust-to-metal}

\begin{figure*}
    \includegraphics[width=0.9\textwidth]{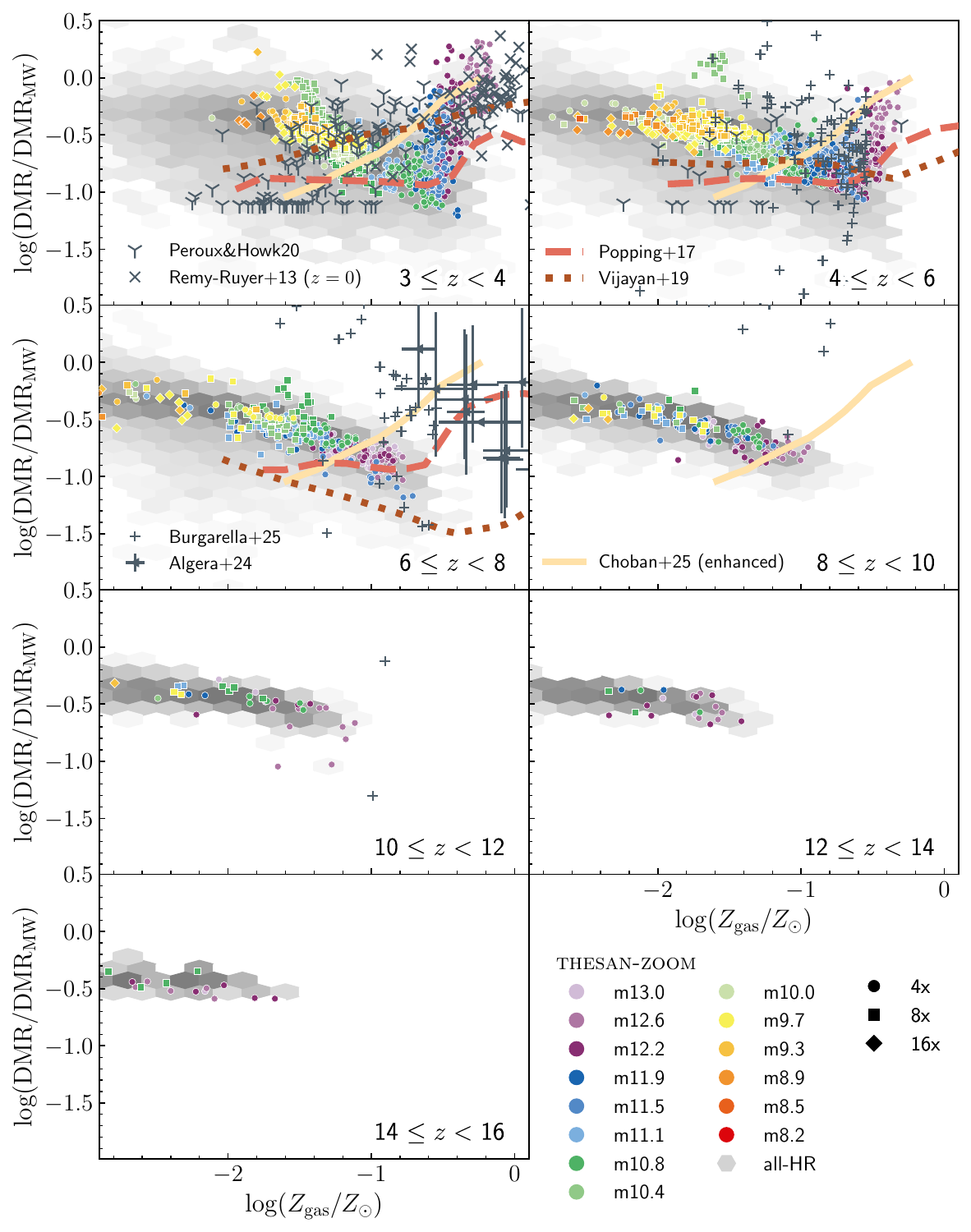}
    \caption{\textbf{Dust-to-metal ratio (DMR) as function of gas-phase metallicity ($Z_\mathrm{gas}$) for \thesanzoom galaxies}. The DMR is normalized by the Milky Way value ($\mathrm{DMR}_\mathrm{MW} = 0.44$). Each panel shows a different redshift range. Colors represent different targets while symbols reflect their resolution level. The gray background histogram shows non-target simulated galaxies. The thick purple contour shows the central 68\% of the data from the original \thesan simulation. Dark grey symbols show observations from \citet[][crosses]{Remy-Ruyer+2013}, 
    \citet[][tripods]{Peroux&Howk2020}, and \citet[plus symbols][]{Burgarella+2025}. The orange dashed, dotted brown and solid beige lines show predictions from the semi-analytical models of \citet{Popping+17} and \citet{Vijayan+2019} and the FIRE-2 simulations of \citet{Choban+2025}, respectively. \textbf{\thesanzoom galaxies predict a V-shaped dependency of the DMR on the gas metallicity, that is not seen in observations.}. }
    \label{fig:DMR_vs_Zgas}
\end{figure*}

Another common relation used to characterize the dust properties of galaxies is the dust-to-metal (mass) ratio (DMR) evolution with gas metallicity. This contains information on the ability of cosmic dust to accrete gas-phase ISM metals (therefore depleting them). We show this quantity in Fig.~\ref{fig:DMR_vs_Zgas}. 

galaxies show an excellent match to the observations from \citet[][crosses]{Remy-Ruyer+2013}. 
\thesanzoom exhibit a sharp inversion of the trend between DMR and gas metallicity at $\log(Z_\mathrm{gas}) \sim 0.5$ unlike the semi-analytical models of \citet{Popping+17} and \citet{Vijayan+2019}. 
The former predicts a constant dust-to-metal ratio at $\log(Z_\mathrm{gas}) \lesssim 0.5$, while the latter is unable to reproduce the large DMR ratios in metal-rich galaxies. Unlike these models, we predict a `bitter spot' for the conversion of metals into dust, where the stellar feedback actively destroys dust injected by stars (see Sec.~\ref{subsec:burstiness_and_dust_survival} for a discussion) but grain growth is not yet efficient. Observations from \citet[][crosses]{Remy-Ruyer+2013} and \citet[][tripods]{Peroux&Howk2020} match our predictions for high-metallicity systems, but do not show such V-shaped dependency at lower gas metallicities. It should be noted that they present a scatter as large as our predicted DMR range and are measured with different techniques, hindering a direct comparison. Therefore, the characterization of the metal and dust content of galaxies in the first billion years of the Universe (or, more generally, in low-metallicity systems) will deliver important clues on the conditions of their ISM, key information to unveil the assembly of the first galaxies. 

Curiously, in this case the observations from \citet{Burgarella+2025} seem to be in approximate agreement with our simulations at $4 \leq z < 6$, although they present a much larger scatter. However, at higher redshifts our galaxies exhibit somewhat lower DMR, although the number of observed galaxies in the same range of gas metallicities is limited. In fact, at $z > 6$ most galaxies with measured DMR are in a metallicity regime where: \textit{(i)} no \thesanzoom galaxy is present, and \textit{(ii)} we expect grain growth to be efficient. Thus, extrapolating the trend seen at $3 \leq z < 4$, it seems likely that the \thesanzoom model would match the bulk of observed galaxies.

\subsection{Dust temperature}
\label{subsec:temperature}

\begin{figure}
    \includegraphics[width=\columnwidth]{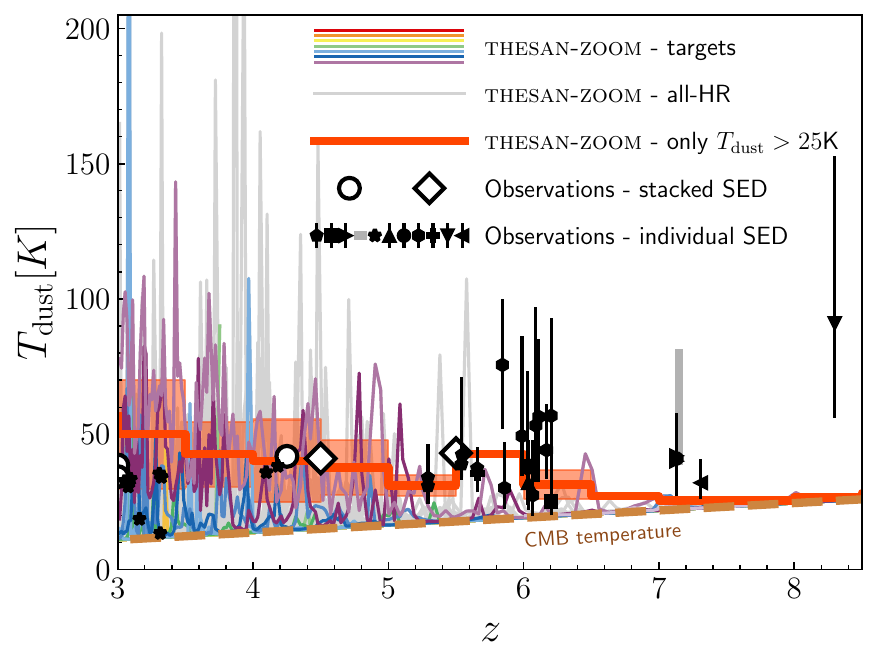}
    \caption{\textbf{Dust temperature evolution} in \thesanzoom. Target galaxies are shown using colored lines, while other high-resolution galaxies identified at $z=3$ are shown in grey. The red histogram shows the average dust temperature and its standard deviation (shaded region) computed assuming only galaxies with $\Tdust>25$\,K are observable, in order to coarsely mimic observational limitations. We also report observations of individual objects from \citet[][pentagons]{Faisst+2020},  \citet[][squares]{Harikane+2020}, \citet[][right-facing triangles]{Bakx+2021}, \citet[][stars]{Bakx+2024a}, \citet[][upward-facing triangle]{Bakx+2024b}, \citet[][downward-facing triangle]{Bakx+2025}, \citet[][grey vertical band, corresponding to different assumptions on the dust emissivity]{Sugahara+2021}, \citet[][circle]{Akins+2022}, \citet[][hexagon]{Mitsuhashi+2024}, \citet[][plus symbol]{Villanueva+2024}, \citet[][left-pointing triangle]{Algera+2024}, as well as measurements derived from stacked SEDs by \citet[][empty circles]{Schreiber+2018} and \citet[][empty diamonds]{Bethermin+2020}. Finally, the brown dashed line indicates the CMB temperature. \textbf{The \thesanzoom galaxies show great variability of $\Tdust$ and can reproduce the observed data once the limited sensitivity of observations is approximately accounted for.}}
    \label{fig:Tdust_evol}
\end{figure}

Now that we have validated our model against commonly-used scaling relations, we move to study the predicted physical properties of cosmic dust in the \thesanzoom galaxies. 
One uniqueness of \thesanzoom is its on-the-fly fully-coupled modeling of cosmic dust, UV and IR radiation, which enables us to self-consistently predict dust temperatures. This offers an important window on the ISM physics. 

In Fig.~\ref{fig:Tdust_evol}, we show the evolution of the dust temperature $\Tdust$ in the \thesanzoom galaxies (colored lines for the target galaxies and grey lines for \candidates). This shows that the dust temperature in our galaxies increases very quickly for short amounts of time. Before discussing the physical origin of this behaviour below, we compare our predicted $\Tdust$ to available observations. 
As discussed in Sec.~\ref{sec:intro}, empirical determination of the dust temperature is difficult, limiting our ability to properly compare our results with data. 
In the Figure we restrict ourselves to observations that sample the IR peak with multiple data points. This makes the estimated dust temperatures more reliable. Namely, we show data from \citet[][pentagons]{Faisst+2020},  \citet[][squares]{Harikane+2020}, \citet[][right-facing triangles]{Bakx+2021}, \citet[][stars]{Bakx+2024a}, \citet[][upward-facing triangle]{Bakx+2024b}, \citet[][downward-facing triangle]{Bakx+2025}, \citet[][grey vertical band, corresponding to different assumptions on the dust emissivity]{Sugahara+2021}, \citet[][circle]{Akins+2022}, \citet[][hexagon]{Mitsuhashi+2024}, \citet[][plus symbol]{Villanueva+2024} and \citet[][left-pointing triangle]{Algera+2024}. Finally, we also plot $\Tdust$ measurements from stacked SEDs by \citet[][empty circles]{Schreiber+2018} and \citet[][empty diamonds]{Bethermin+2020}. 

As anticipated, the average dust temperature in \thesanzoom galaxies remains below the observed value for most of their lifetimes. However, when the dust is heated, the temperature easily reaches values comparable to or larger than those observed. In fact, for each observed data point we can find one simulated galaxy with the same $\Tdust$ within a redshift window $\Delta z \approx 0.5$. This is remarkable given the somewhat limited number of simulations in the current \thesanzoom suite. 
The only exceptions are the observed dust temperatures in A1689-zD1 \citep{Bakx+2021, Sugahara+2021} at $z = 7.13$ (notice that all data points at $z=7.13$ are different observations of the same galaxy) and in  MACS0416\_Y1 \citep{Bakx+2025} at $z=8.13$. However, this is likely a consequence of our limited galaxy sample, since the estimated stellar masses of these objects are $\Mstar \sim 10^9 \, \Msun$ \citep[][]{Watson+2015, Bakx+2025}, at the upper limit of the simulated \thesanzoom sample at $z=7$ and, and therefore very poorly sampled by our simulation suite. We also note that the majority of these galaxies are significantly more massive than most of our simulated objects, and therefore likely present more extreme physical conditions.

In order to better compare to observed galaxy samples, we approximately account for the finite observational sensitivity by assuming that only galaxies with $\Tdust>25$K can be observed.\footnote{This is a very simplistic approach, since each observation has varying depth and the minimum observable dust temperature depends on the dust content and redshift of the galaxy. However, a detailed forward modeling of our galaxies to enable faithful comparison with observations is beyond the scope of this paper and will be presented in an upcoming work (Popovic et al., in prep.).} 
This temperature roughly corresponds to the lowest dust temperature that has been reliably constrained in the $z\gtrsim4$ galaxies (see observations in the Figure). We have also checked that the exact temperature threshold has a negligible impact on the results as long as it is above the CMB temperature. 
Applying this selection to the galaxy population to determine the `observable' galaxies, we find that the average temperature (red histogram in the plot; the shaded area around it corresponds to the standard deviation within each redshift bin) in \thesanzoom is very close to the observed one in galaxies with reliable temperature estimations. 
Nevertheless, it should be noted that observational samples, although heterogeneous in their properties, effective volume and sensitivity limits, seem to indicate that the average dust temperature in galaxies increases with redshift. This trend is particularly visible at $z \lesssim 4$ \citep[\eg][]{Bethermin+2015, Schreiber+2018, Viero+2022}. On the other hand, the \thesanzoom galaxies tend to have higher average dust temperatures at \textit{lower} redshift. If this trend continues significantly past the final simulation redshift (\ie at $z < 3$), this would put them in disagreement with observations. Further work extending our runs to lower redshift is needed to finally assess whether such disagreement exists. However, as we discuss in Sec.~\ref{sec:discussion}, this is expected as a consequence of the burstiness of our model. 

Another noteworthy feature seen in Fig.~\ref{fig:Tdust_evol} is the fact that, especially at lower redshifts, our simulations predict dust temperatures in excess of $100 K$, significantly above the observed ones. While some of these are found in small objects, which could therefore be difficult to observe, we also find such large values in the most massive galaxies in our sample. This rises concerns, as extreme temperatures like these should be easily observable but have not been detected, thus pointing towards a shortcoming of our model. This is once again related to the bursty nature of star formation as described above. However, as shown below, even matching the SFR of our galaxies to observations still results in simulated dust temperatures in excess of observations. A possible explanation of such behaviour is the lack of very dense clumps because of limited resolution. These clumps could more effectively shield most of the dust from the UV radiation of newborn stars, lowering the average temperature. This effect could be exacerbated by our feedback injection scheme, which distributes the energy over a kernel rather than injecting it directly on the mesh faces. This implies that, even if such small clumps exist, they could be effectively bypassed by the injection scheme if they are fully within the injection kernel. It should be noted, however, that the dust emission is strongly dominated by the hot dust component, thus it is not clear how effective such mechanism could be. 

\begin{figure}
    \includegraphics[width=\columnwidth]{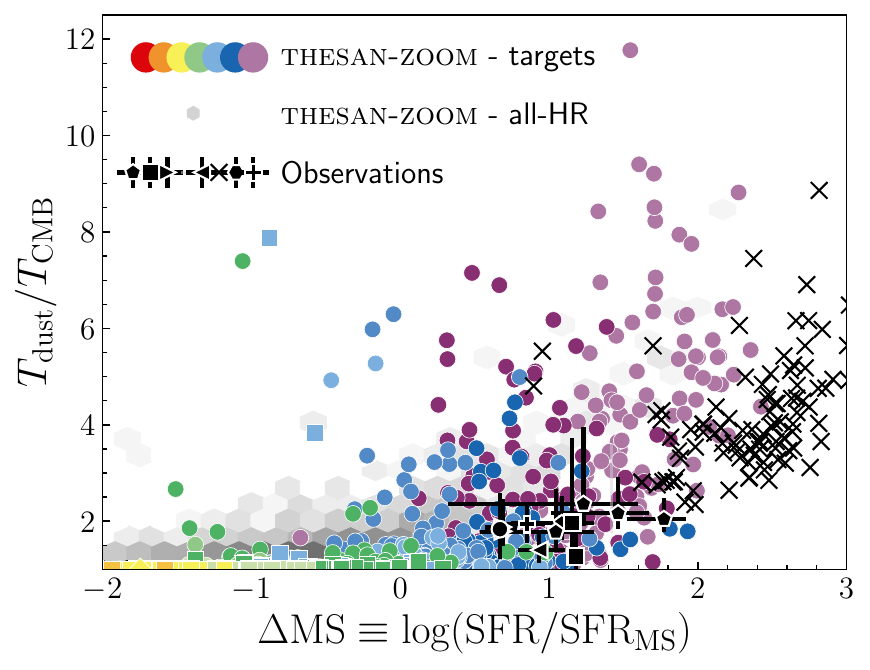}
    \caption{\textbf{Impact of SFR on dust temperature}. The Figure shows the dust temperature excess over the CMB ($T_\mathrm{dust}/T_\mathrm{CMB}$) as a function of distance from the star-forming main sequence (MS), \ie $\Delta \mathrm{MS} \equiv \log(\mathrm{SFR}/\mathrm{SFR}_\mathrm{MS}$. We use a 10 Myr-averaged SFR for our galaxies and the MS definition in \citet{ThZoom_bursty}. We show all available outputs for \thesanzoom target galaxies (colored circles) as well for other \candidates (grey background histogram). We also show measurements from \citet[][pentagons]{Faisst+2020},  \citet[][squares]{Harikane+2020}, \citet[][right-facing triangles; stellar mass and SFR taken from \citealt{Bradley+2008}]{Bakx+2021}, \citet[][grey vertical band corresponding to different assumptions on the dust emissivity; the stellar mass is taken from \citealt{Hashimoto+2019}]{Sugahara+2021}, \citet[][circles; stellar mass and SFR taken from \citealt{Bradley+2008}]{Akins+2022}, \citet[][hexagon]{Mitsuhashi+2024}, \citet[][plus symbol]{Villanueva+2024} and \citet[][left-pointing triangle]{Algera+2024}. We also show measurements from \citet[][]{daCunha2015} because they span a large range of $\Delta \mathrm{MS}$, although they do not necessarily sample the IR dust peak with multiple data points, and therefore should be considered less reliable than the others in this plot \citep{Bakx+2021, SommovigoAlgera2025}. For all data we use the same definition of main sequence as for our simulations. 
    \textbf{Galaxies above the main sequence feature hot dust because of the excess of UV photons produced by young stars within the galaxy.}
    }
    \label{fig:Tdust_DMS}
\end{figure}

As mentioned above, the rise in dust temperature is driven by starbursts. We explicitly show this quantitatively in Fig.~\ref{fig:Tdust_DMS}, where we plot the dust temperature excess over the CMB as a function of distance from the star-forming galaxy main sequence (MS) $\Delta \mathrm{MS} \equiv \log(\mathrm{SFR}/\mathrm{SFR}_\mathrm{MS}$. We use a 10 Myr-averaged SFR for our galaxies and the MS definition in \citet{ThZoom_bursty}. Since we do not expect redshift to play a role in this relation (beyond setting the MS), we show all available outputs together. The figure clearly shows that as long as our galaxies lie close to or below the MS, their dust remains at the floor imposed by the CMB temperature. However, as soon as the star formation rate is a few times larger than the MS, the dust is heated up significantly by the UV photons produced by newborn stars. 
Shortly after the beginning of the starburst, the dust temperature drops very quickly to the CMB floor. In fact, this typically occurs \textit{before} the end of a starburst episode (see also Fig.~\ref{fig:tracks} and related discussion). The reason for such a quick and sharp drop in our model is two-fold. Initially, radiation pressure, stellar winds, and our early stellar feedback push dust away from the star formation locations, reducing its exposure to the intense UV radiation of newborn stars and therefore allowing it to cool despite the strong UV field from newborn stars. Then, once stars begin exploding as SN, they efficiently destroy dust in their surroundings through sputtering, preferentially removing hot dust and, therefore, reducing the average dust temperature in a galaxy. 
This picture starts to break down for the most massive galaxies in our sample. \texttt{m12.6} and \texttt{m12.2} never cool down within 10\% of the CMB temperature after, respectively, $z \lesssim 5.4$ and $z\lesssim4.5$ (see Fig.~\ref{fig:Tdust_evol}). While the $\Tdust$ evolution remains bursty, new stars are formed sufficiently often to maintain the dust  hotter than the CMB. This suggests that even more massive systems would be able to sustain a warm/hot dust content in our model. We have quantified the duration of hot-dust phases in our galaxies by identifying local maxima in $\tilde{T}(z) \equiv T_\mathrm{dust}(z)-T_\mathrm{CMB}(z)$ and then define the duration of such phase as the length of the continuous set of redshifts around the peak having $\tilde{T}(z) \geq 0.1 \tilde{T}_\mathrm{local\,max}$. We find a median duration of $20.3_{\mathsmaller{-2.4}}^{\mathsmaller{+2.3}}$ Myr.

As a consequence of the dust life cycle described above, our model predicts that observed IR-bright galaxies are just a small fraction of the total and are experiencing an intense starburst episode. This also implies that their properties are likely biased compared to the bulk of the population, and underlines the importance of deep IR observational campaigns able to measure the properties of galaxies outside of such extreme phases of their lives. 

We compare our prediction with data from \citet[][pentagons]{Faisst+2020},  \citet[][squares]{Harikane+2020}, \citet[][right-facing triangles; stellar mass and SFR taken from \citealt{Bradley+2008}]{Bakx+2021}, \citet[][grey vertical band corresponding to different assumptions on the dust emissivity; the stellar mass is taken from \citealt{Hashimoto+2019}]{Sugahara+2021}, \citet[][circles; stellar mass and SFR taken from \citealt{Bradley+2008}]{Akins+2022}, \citet[][hexagon]{Mitsuhashi+2024}, \citet[][plus symbol]{Villanueva+2024},  \citet[][left-pointing triangle]{Algera+2024} and \citet[][]{daCunha2015}. For the last one, we caution the reader that these are based on observations that do not constrain well the IR dust. Thus, as shown in \citet{Bakx+2021, SommovigoAlgera2025}, they are likely affected by large uncertainties and/or systematic shifts. We choose to show them nevertheless because they span a large range of $\Delta \mathrm{MS}$, and therefore provide a more comprehensive view of observational results. 
We find that data are in overall agreement with our prediction, although dust in our simulated galaxies seems to be somewhat hotter than observed. However, considering the large scatter in both the simulated and observed relations, the dependence on the exact definition of main sequence and the limited sample size, we find this level of agreement satisfactory. 

\subsection{Radial profiles}
\label{subsec:profiles}

\begin{figure}
    \includegraphics[width=\columnwidth]{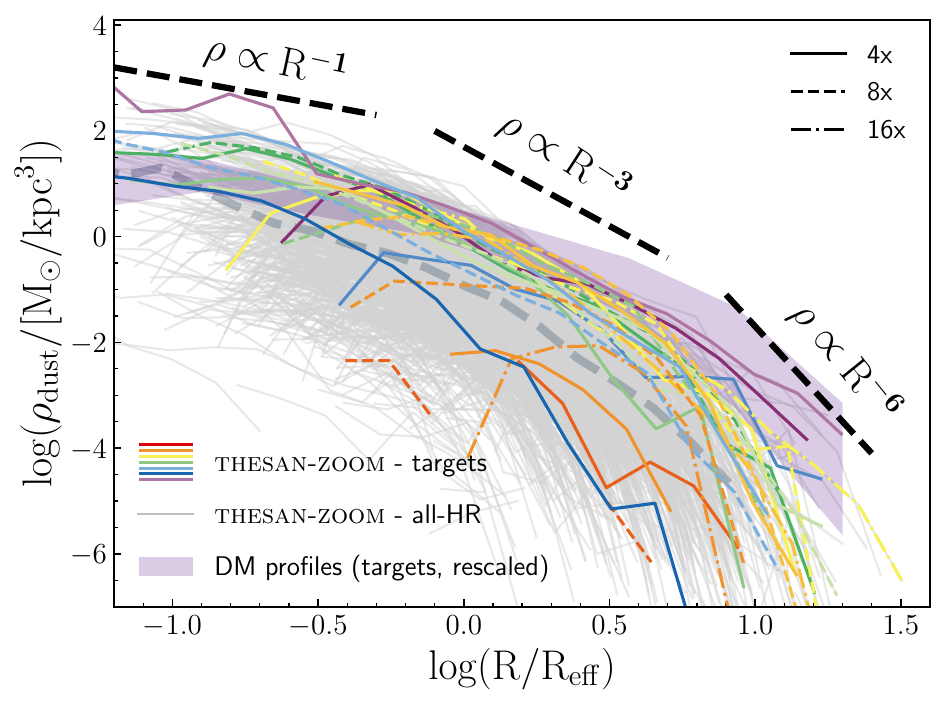}
    \caption{\textbf{Radial dust density profile} at $z=3$ for \thesanzoom targets (colored lines) and \candidates (grey thin lines, their median profile is indicated by a dark grey dashed line). Different zoom factors are reported with different line styles, as indicated in the legend. The radial distance is normalised by the effective radius, defined as twice the stellar half mass radius. For reference, we show with dashed lines $R^{-1}$, $R^{-3}$ and $R^{-6}$ slopes. The purple shaded region shows the central 68\% distribution of the targets' dark matter density profile, rescaled by their global dust-to-dark matter mass ratio. \textbf{The dust distribution within \thesanzoom galaxies approximately follows an NFW profile in the center but drops off more steeply at larger radii}.} 
    \label{fig:dust_profiles}
\end{figure}

The last property that we investigate in this Section is the distribution of dust within simulated galaxies. In Fig.~\ref{fig:dust_profiles}, we show the spherically-averaged radial dust profiles for the \thesanzoom galaxies at $z=3$, with distance normalised by the stellar half-mass radius of each galaxy. 
We show the target galaxies using colored lines, whose style identifies the zoom level (as indicated in the plot legend) and the \candidates are shown in grey. 
The profiles are approximately described by an NFW profile \citep{NFW}, although more concentrated than for the stellar component. For reference, we show with a purple shaded area in the Figure the central 68\% of the targets' dark matter density profiles, each rescaled by the dust-to-dark matter ratio of the galaxy. These are generally in good agreement with the dust ones in the outer part of the galaxy. Many simulated galaxies show signs of dust depletion. This is limited to the central regions for the largest objects in our sample but extends to most or all of the galaxy for the smaller ones. 
As discussed above, this is a consequence of the strong feedback during starburst events. 

\begin{figure}
    \includegraphics[width=\columnwidth]{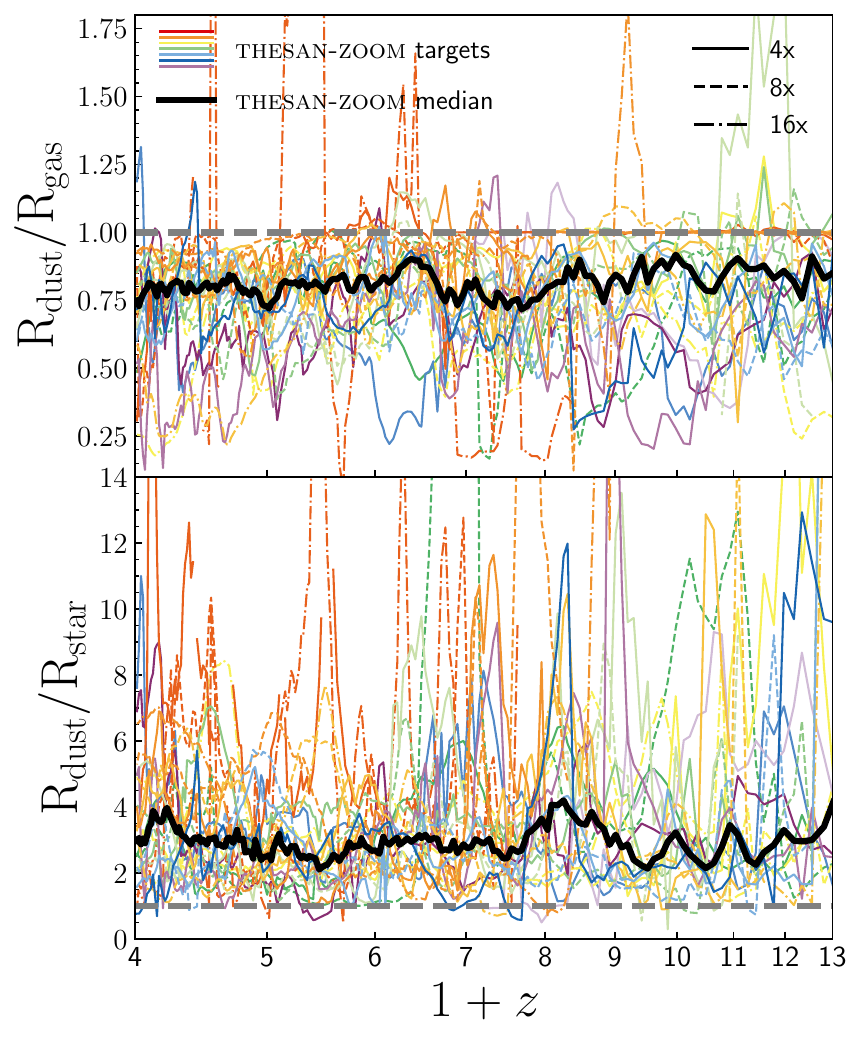}
    \caption{\textbf{Dust-to-gas (top) and gas-to-stellar (bottom) size evolution}. All radii are computed as twice the half-mass radius of the component, while the thick black lines show their median. Thin lines show the value for individual \thesanzoom targets, with different line styles corresponding to different zoom factors. We do not show other \candidates for visual clarity. \textbf{The dust distribution is typically more concentrated than gas and more extended than stars, and experiences strong variations over short time scales.}}
    \label{fig:dust_radius}
\end{figure}

Although the overall picture does not evolve much with redshift, individual profiles change dramatically with time. To show this, in Fig. \ref{fig:dust_radius} we plot the ratio between the dust and gas radii (top panel) or stellar radius (bottom panel) as a function of redshift. All radii are computed as the half-mass radius of the component. 
In the Figure, each colored line represents a \thesanzoom target (we omit the \candidates for visual clarity) and the black solid line shows the median evolution. All galaxies show quick variations in the relative spatial extents of dust, gas, and stars.  Nevertheless, some general features are also clearly visible. On average, the dust distribution is more compact than that of the gas at all redshifts. This owes to dust being formed primarily in high-density regions close to the center of the halo potential well. 
However, while strong feedback episodes preferentially destroy the dust closer to the halo center, they also efficiently expel the gas, so that only very rarely $\mathrm{R}_\mathrm{dust} > \mathrm{R}_\mathrm{gas}$. 

Since the stellar distribution is not affected by feedback episodes, the bottom panel of Fig. \ref{fig:dust_radius} offers a better view of the change in the dust distribution during those phases. $\mathrm{R}_\mathrm{dust}$ can vary up to an order of magnitude over relatively short timescales.

\section{Observed properties of thesan-zoom galaxies}
\label{sec:observed_properties}

The simulated properties analysed in the previous Section are not directly accessible by observations. For this reason, in this Section we forward model our galaxies to provide more direct comparisons with available data. 

\subsection{UV--IR offset}
\label{sec:offsets}

\begin{figure}
    \hspace*{-0.05\columnwidth}%
    \includegraphics[width=1.05\columnwidth]{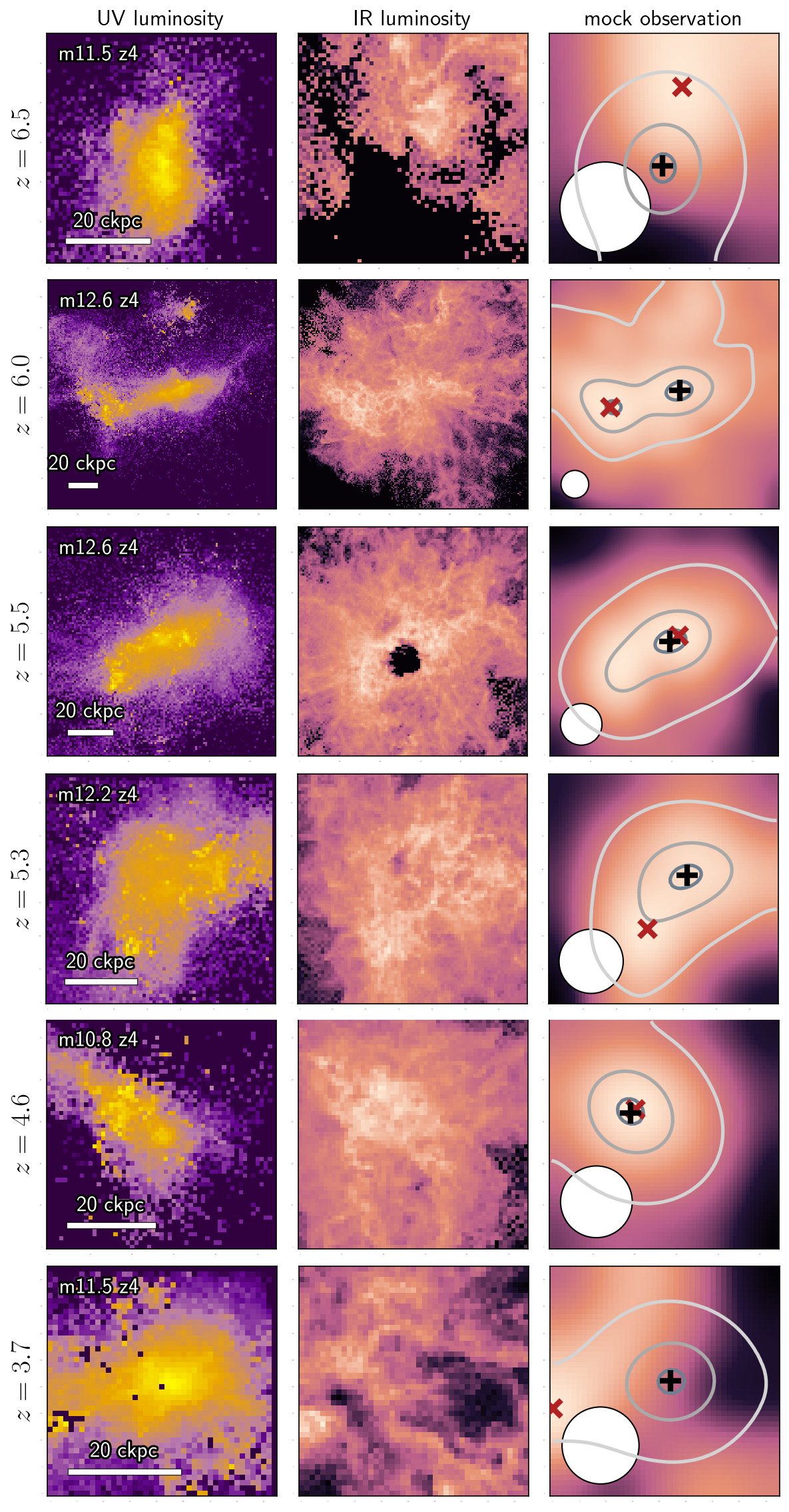}
    \caption{\textbf{Examples of UV--IR offsets} (or lack thereof) in simulated galaxies at $z=6$. Each 3-panel collage refers to a different redshift (reported on the left) and galaxy (reported on the top left, followed by the zoom level). Within each collage, the left and central panels show the projected attenuated UV luminosity in arbitrary units and the IR luminosity as traced by the projected total dust mass. In the right panel, we show both luminosities (UV as grey contours, IR as background) smoothed using a Gaussian filter with $\sigma = 0.5$ physical kpc, to approximately mimic the beam size of observations (shown as a white circle in the bottom left of the panel). The red cross and black plus indicate the maxima of the IR and UV luminosity distributions, respectively. The third row from the top shows a case where the offset separation is smaller than half the beam size, and therefore is excluded from the analysis (see text). Notice that each panel has a different physical size, corresponding to four times the halo stellar half-mass radius. \textbf{The \thesanzoom galaxies feature a variety of dust and stellar morphologies, creating a distribution of UV--IR offsets.}}
    \label{fig:offset_maps}
\end{figure}

\begin{figure*}
    \includegraphics[width=\textwidth]{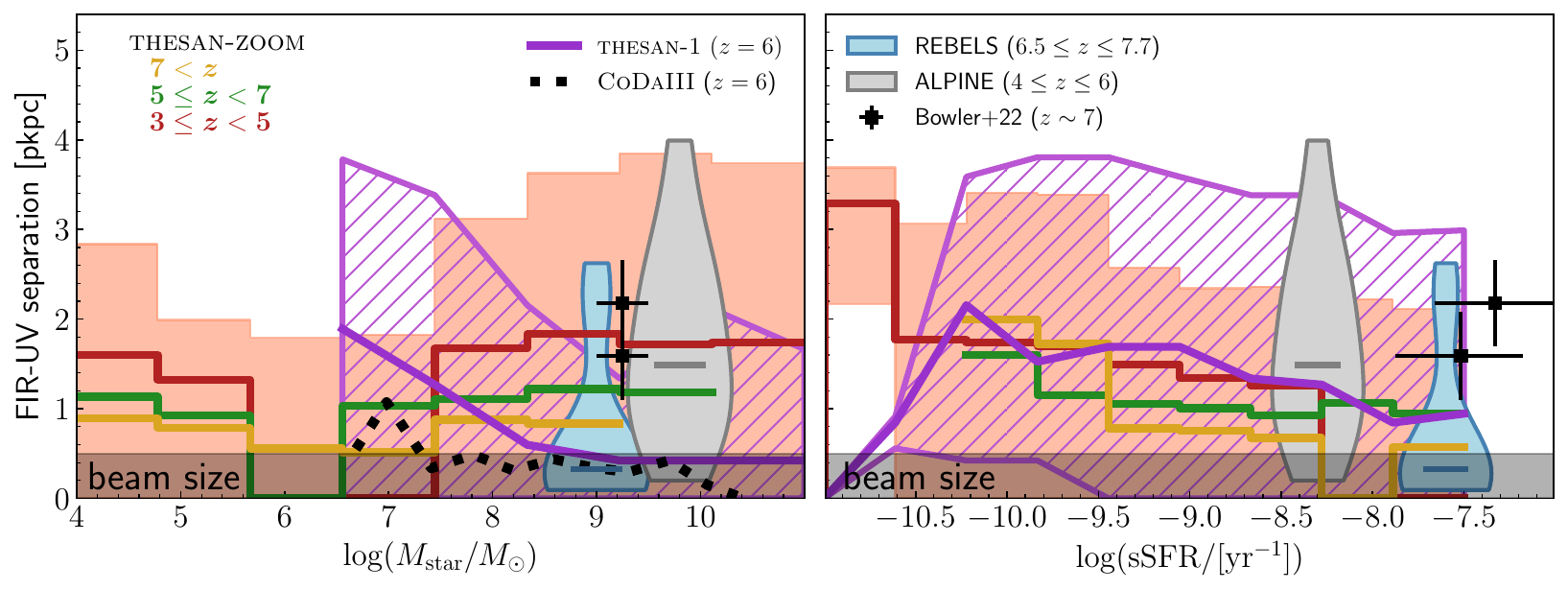}
    \caption{\textbf{UV--IR spatial offsets} as function of stellar mass (left) and specific star formation rate (right). The median for \thesanzoom galaxies in three redshift bins (red, green and yellow histograms, see legend) is shown using a solid histogram, while the central 68\% of the data is shown by a shaded region around it. For visual clarity, we only show the latter for the central redshift bin, but it has a similar extent in all bins. The greyed band at the bottom shows the (assumed) beam size (see text). We also show the results from the original \thesan box (purple histogram, hatched region) and from the \texttt{CoDa III} simulation \citep{Ocvirk+25}. Finally, we report the distribution of observational results from the REBELS \citep[blue violin,][]{rebels_offset} and ALPINE \citep[grey violin,][]{alpine_offsets} surveys, including their median (horizontal bar), and the results from \citet[][black squares]{Bowler+2022}. The horizontal location of the violin corresponds to the approximate average stellar mass and sSFR from all galaxies in the observed sample. \textbf{\thesanzoom predicts UV--IR offsets in very good agreement with the observed ones, approximately constant with host galaxy stellar mass and decreasing with increasing sSFR}.}
    \label{fig:offset_vs_Mstar_sSFR}
\end{figure*}

Multi-wavelength observations of dusty star-forming galaxies revealed the widespread existence of spatial offsets between the peak of UV/optical emission tracing the (young) stellar component, and the IR emission from cosmic dust \citep{Goldader+2002, Bowler+2022, rebels_offset, alpine_offsets, cristal, Spilker+2023, Liu+2024, Bakx+2025} for individual objects or small samples. The physical origin of such offsets is still debated. Broadly speaking, there are two main possibilities, namely that the dust and stellar components are spatially distinct, or that the dust attenuation effectively hides the bulk of UV light emitted by young stars, allowing only the detection of secondary peaks in the UV distribution. Here, we investigate this question using our simulations. 

In order to provide a faithful comparison with observations, we construct mock observations of our simulated galaxies. We start by making the simplifying assumption that all relevant processes (\ie UV/IR emission and dust absorption) occur within 10 stellar half mass radii\footnote{We note that, while this might miss some extreme cases (see \eg Fig.~\ref{fig:dust_radius}), it is not expected to significantly bias our statistical sample as such cases are rare and would most likely be identified in observations as well.} ($R_\mathrm{shm}$) of the galaxy \citep[see also][]{Ocvirk+25}. Therefore, we extract data cubes centered on the most bound gas cell in the galaxy and extending $10 R_\mathrm{shm}$ in each positive and negative direction. 
Then, we interpolate particle information on a Cartesian grid using a cloud-in-cell approach. Using each stellar particle as a source, we perform ray tracing along three perpendicular directions to obtain surface density maps of attenuated UV emission. We use the BPASS library to estimate the intrinsic UV emission at $1500$ \AA\xspace of each stellar particle. We assume the IR emission perfectly traces the projected dust mass, as done in \citet{Ocvirk+25}. 
To account for dust attenuation, we leverage the fact that our simulations predict the spatially- and temporally-varying Carbon and Silicon content of dust to compute the attenuation due to the exact mix of carbonaceous and silicate dust using the model of \citet[][]{Draine&Lee1984, Laor&Draine1993}\footnote{The optical properties are available at \url{https://www.astro.princeton.edu/~draine/dust/dust.diel.html}}, rather than assuming \eg a SMC dust mix as done in previous similar studies \citep[\eg][]{Ocvirk+25}. However, for reasons discussed below, we find that assuming an SMC-like dust mix does not significantly change our results. As in \eg \citet{Ocvirk+25}, we account for absorption only and not for dust albedo. \citet{Ocvirk+25} argue that, because of the isotropic nature of stellar sources and the approximately spherical distribution of gas and dust, this has no impact on the result. In a forthcoming paper (Popovich \textit{et al.}, in prep.) we will analyse synthetic Spectral Energy Distributions of our galaxies produced by the radiative transfer \texttt{SKIRT} code \citep{skirt} and, therefore, we will be able to quantify the impact of this assumption more accurately. 
Finally, we convolve the 2D maps with a point spread function (PSF). Since each observed galaxy has a different PSF and we are interested in a population-level comparison, we approximate the PSF using a Gaussian with $\sigma=0.5$ kpc (physical), representative of the average beam width of the observational dataset we use for comparison. 

We show in Fig.~\ref{fig:offset_maps} some examples of the outcome of the procedure described above. Each row shows a different galaxy (indicated in the top left corner of the panels) at a different redshift (reported on the left). 
Columns from left to right show the projected UV luminosity, the projected attenuated IR luminosity and a `mock observation' where the UV (contours) and IR (background image) luminosities have been convolved with the aforementioned PSF (indicated by the white circle in the bottom left of the panel). We mark with a black plus symbol and a red cross the locations of the UV and IR maxima, respectively. Notice that, because of the selection procedure outlined above, each panel covers a different physical scale, as shown by the ruler in the bottom left of the leftmost panels. 

A striking first feature emerging from our simulations is the ubiquity of morphological differences between the UV and IR light distributions. For example, the top row of the Figure shows how almost all the dust in the depicted galaxy lies in regions with little UV radiation. It is interesting to note that the peak of the UV distribution corresponds to regions with no dust at all. This results in a clear, detectable UV--IR offset (see the corresponding right panel). A similar, although less-extreme, configuration is shown in the bottom row. Another interesting case is shown in the third row from the top, where the IR distribution shows a clear hole, which however does not coincide with any bright UV peak. Interestingly, when observational resolution is taken into account, such a hole is hidden and there is essentially no UV--IR offset. In the other cases shown, the UV and IR emission visually trace each other well. Nevertheless, when such distributions present multiple peaks, it is common to find significant offsets originating from UV and IR light that dominate different peaks. 

In order to quantitatively compare our measurements to available datasets, we systematically compute the distance between the global maxima of the UV and IR distributions in the (PSF-convolved) synthetic images. We reject cases where the local maxima identified are within half the beam size, resulting in approximately 9500 offset values. Fig.~\ref{fig:offset_vs_Mstar_sSFR} shows their distribution as a function of stellar mass (left) and specific star formation rate (sSFR, right) of the galaxy. The colored histograms show the distribution of offsets in three different redshift bins, with the solid line indicating the mean and the shaded region around marking the central 68\% of the data. The latter is very similar for all redshift bins, so we show it only for one histogram, for the sake of visual clarity. The median offset is remarkably constant across galaxies with different stellar masses, and only in the highest-redshift bins does it show a hint of positive correlation. However, we find hints of a negative correlation of the UV--IR offset with the specific star formation rate of the galaxy, but limited to low-sSFR galaxies.

We compare our results with the original \thesan simulation (purple line and hatched region around it), where we show only the results at $z=6$ for visual clarity. In this case, the offsets are computed using the same procedure described above. The trend with sSFR (right panel) is consistent with the results from \thesanzoom, but in \thesan the offsets are systematically smaller for galaxies with large stellar masses ($M_\mathrm{star} \gtrsim 10^{8.5}\, \Msun$) and show a sharp increase for smaller galaxies. Although we only include galaxies with at least 10 stellar particles and 100 gas particles, it is possible that resolution effects are driving such an increase. An in-depth study of the original \thesan results is beyond the scope of this paper. When looking at the trend with sSFR (right panel), instead, our simulations are in excellent agreement with the original \thesan simulation. 
In the left panel of the Figure we also show the results from the CoDa III radiation-hydrodynamical simulation \citep{CoDaIII, Ocvirk+25}. Note that CoDa III has a much lower resolution than both our simulations and \thesan within the galaxies, as a consequence of their use of a uniform Cartesian grid across the entire simulation volume. The CoDa III results are in broad agreement with those of the original \thesan simulation, and thus significantly below those of our estimates. They also find an increase in the UV--IR offset towards smaller stellar masses, although much milder than in \thesan. 

Finally, we also show results from the ALPINE \citep[grey violin, ][]{alpine_offsets} and REBELS \citep[blue violin, ][]{rebels_offset} ALMA observational campaigns. For each of them, we report the median offset with a horizontal bar. The horizontal position of the violins is chosen to be the median inferred stellar mass (left panel) or sSFR (right panel) of the sample. We use black squares to show the observations from \citet{Bowler+2022}. 
Our results are in good agreement with the ALPINE measurement, while we find a mild tension with the REBELS measurements when matching the stellar masses, as \thesanzoom predicts larger offsets. It should be noted, however, that estimating stellar masses of these galaxies is notoriously difficult. In fact, taking the REBELS and ALPINE data at face value would imply a rapid evolution of the offsets with stellar mass and redshift, in tension with current models.

\subsubsection{Origin of UV--IR offsets}
To understand the origin of the UV--IR offsets, we compare the intrinsic and dust-extincted UV maps. We produce the former by repeating the procedure described above to produce synthetic maps but do not attenuate the UV luminosities. Then, we compute the offset between the peaks in these two maps. If dust extinction plays a relevant role in creating these offsets by hiding UV bright spots, we expect these intrinsic-extincted offsets to be comparable in magnitude and number to the UV--IR offsets. However, we find that in only less than 3\% (0.5\%) of the cases these offsets are larger than 0.1 kpc (1 kpc). Thus, we conclude that dust attenuation does not play a role in creating these offsets. 

Instead, in our simulations it is the physical removal of dust from UV-bright regions, due to feedback from stellar evolution and supernova explosions, that creates UV--IR offsets (see Fig.~\ref{fig:offset_maps} and Fig.~\ref{fig:tracks} and related discussion).

\subsection{Dust surface density distribution function}
\label{sec:dsddf}

\begin{figure}
    \includegraphics[width=\columnwidth]{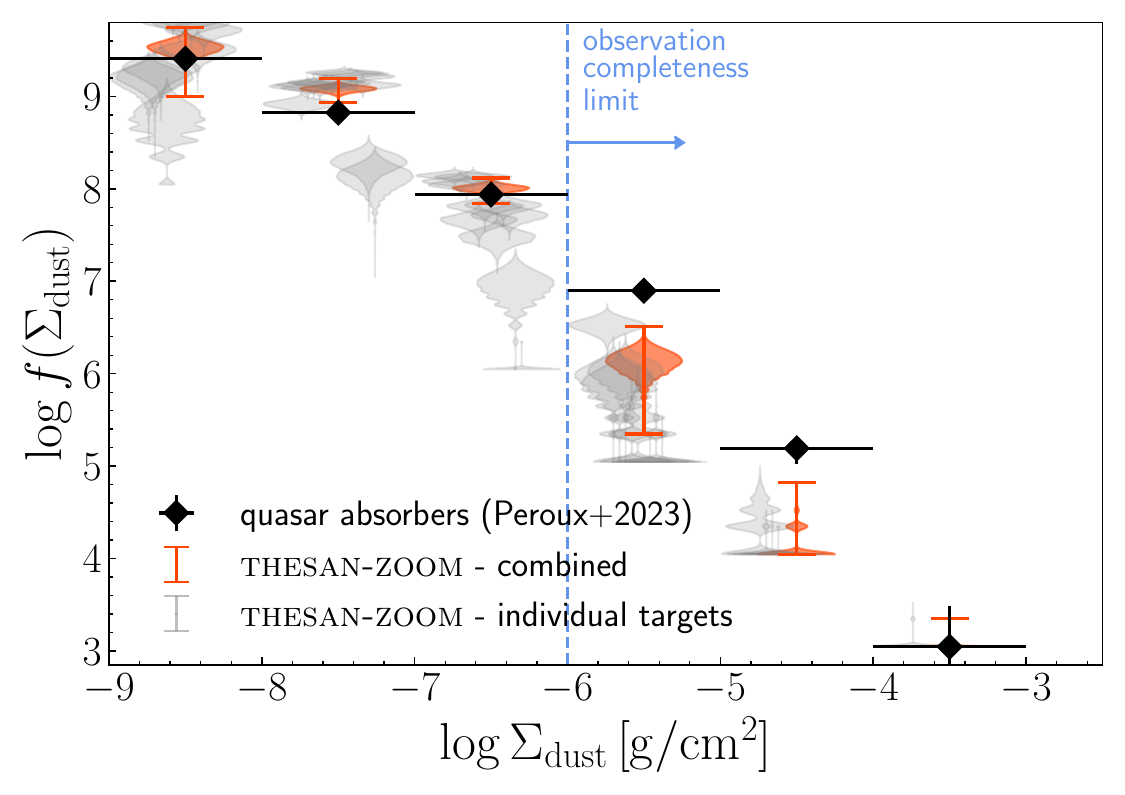}
    \caption{\textbf{Dust surface density distribution function} $f(\Sigma_\mathrm{dust})$ in \thesanzoom galaxies at $z=6$. Light-grey violins show the distribution from $10^4$ random sampling of sightlines through each of the target galaxies, while the red violin shows the distribution of the full sample. For the latter, we mark the extremes of the distribution with horizontal bars. We show the observations from \citet{Peroux+2023} using black diamonds; we only use their quasar sightlines samples (see text). The blue dashed vertical line shows the observational completeness limit inferred by \citet{Peroux+2023} based on the break in the power-law scaling. \textbf{\thesanzoom reproduces the low-$\Sigma_\mathrm{dust}$ part of the distribution well, but it underpredicts the incidence of large dust column densities.}}
    \label{fig:dsddf}
\end{figure}

Observing dust in emission is notoriously difficult and potentially subject to large biases due to the degeneracy between dust mass, temperature, and emissivity \citep[see \eg][]{Bakx+2021}. An alternative approach has been proposed in \citet{Jenkins2009}, where the dust content of galaxies was measured through absorption features in the spectrum of background quasars. This technique allows a measurement of the dust surface density, $\Sigma_\mathrm{dust}$, without any assumptions  about the total gas+dust metallicity. \citet{deCia+2016, Peroux+2023} applied such an approach to Damped Lyman-$\alpha$ systems at $z>2$ identified in quasar spectra to derive a dust surface density distribution function $f(\Sigma_\mathrm{dust}) \equiv \mathcal{N}/\Delta\Sigma_\mathrm{dust}$ (where $\mathcal{N}$ is the number of absorbers in the dust surface density bin $\Delta\Sigma_\mathrm{dust}$; Note that this function is not normalised by the redshift path, and thus depends on the number of sightlines in the survey). 
It should also be noted that observations at low-$\Sigma_\mathrm{dust}$ might be incomplete. The vertical dashed blue line in the Figure corresponds to the completeness limit of the observations reported by \citet{Peroux+2023}, inferred from the break in the power-law scaling of $f(\Sigma_\mathrm{dust})$.

In order to compare faithfully with these kinds of observations, we first estimate the dust surface density of each target galaxy in \thesanzoom at $z=6$ by depositing the dust mass density onto a Cartesian grid using a Cloud-in-Cell algorithm. The cell length is $l_\mathrm{cell} = 0.1$ kpc. For each galaxy we create three projections (along the main axes of the box), which we treat as a collection of sightlines intercepting the galaxy at different locations. Then, we randomly sample from the entire pool (combining all sightlines from all galaxies) the same number of dust surface density values as the number of sightlines in \citet{Peroux+2023}\footnote{Note that the impact parameter distribution is not available in observations, as these systems are identified as damped Lyman-$\alpha$ systems in the quasar spectrum. See Section~6 of \citet{deCia+2016} for a discussion of potential biases introduced by this selection.}. We repeat this step 10\,000 times and show the distribution of predicted values using red violins in Fig.~\ref{fig:dsddf}. In each violin, the top and bottom horizontal bars show the edge of the distribution. To depict the object-to-object variation, we also show using light grey violins (slightly offset from each other in the horizontal direction for visual clarity) the distribution of dust surface densities extracted from single simulated galaxies. Finally, we report the observations from \citet{Peroux+2023} using black diamonds, with horizontal bars indicating the width of the bin used (identical for these authors and for us) to compute the distribution function. 

Our simulated galaxies reproduce the low-$\Sigma_\mathrm{dust}$ part of the observed distribution well, but struggle to match the high-$\Sigma_\mathrm{dust}$ tail of the distribution. The reason might simply lie in the lack of large systems due to our limited statistics. We will soon test this using new \thesanzoom runs focusing on more massive galaxies. Another possible explanation is lack of sufficient numerical resolution to capture dense systems within galaxies. However, we find no trend in $f(\Sigma_\mathrm{dust})$ when comparing the same objects at different resolution level (although this is available only for small objects that do not contribute to the high-$\Sigma_\mathrm{dust}$ part of the distribution). Considering the discussion in Sec.~\ref{subsec:dust_mass} and Sec.~\ref{subsec:attenuation}, we favor an alternative interpretation, namely that the intrinsic burstiness of our galaxy formation model, which allows us to match the observed UV luminosity function at $z \gtrsim 9$, prevents the build up of large amounts of dust in the \thesanzoom galaxies. We discuss this more in Sec.~\ref{sec:discussion}.

\subsection{Dust attenuation}
\label{subsec:attenuation}

\begin{figure}
    \includegraphics[width=\columnwidth]{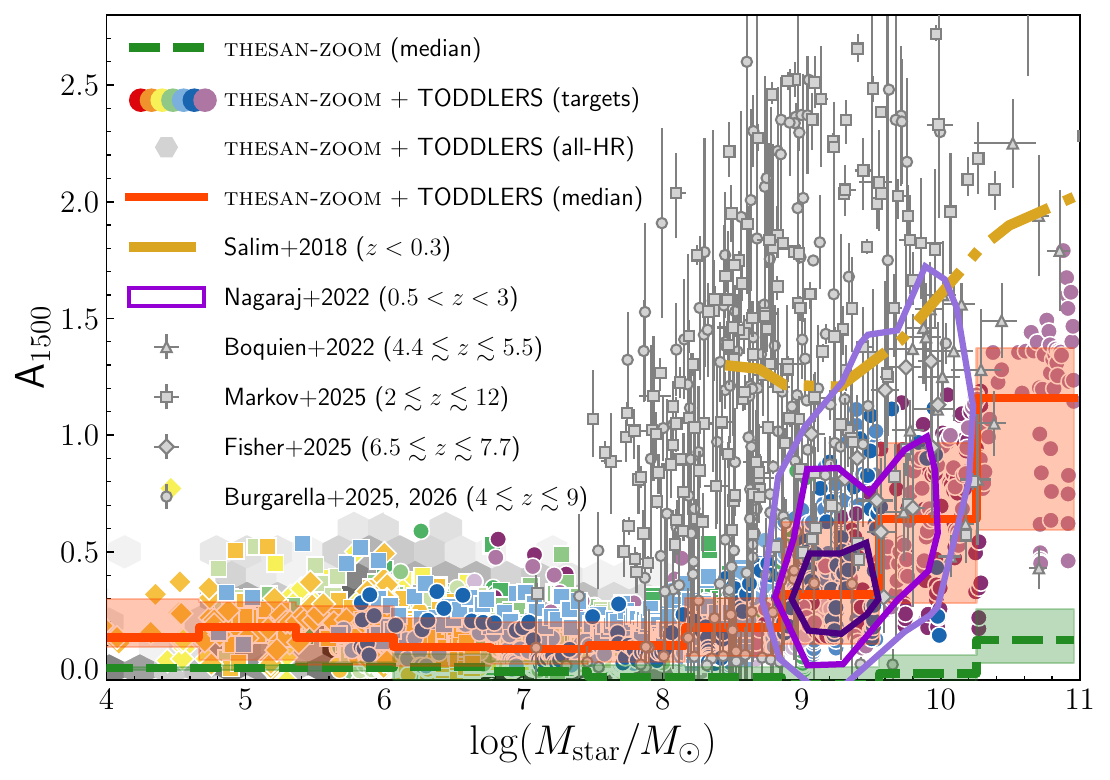}
    \caption{\textbf{Attenuation at 1500\,\AA~as a function of stellar mass} in the \thesanzoom galaxies, augmented with the TODDLERS model for unresolved dust in the stellar birth cloud. The colored symbols show the target galaxies (using circles, squares and diamonds for zoom factor 4, 8, and 16, respectively). The background gray shaded region shows the distribution of \candidates at integer redshifts only. The histogram and the surrounding shaded region show the median and central 84\% of the data in bins of stellar mass. We report the results for the native \thesanzoom (\ie without the TODDLERS model) with a green dashed histogram. We also show observations at different redshifts from \citet[][black squares]{Markov+2025a,Markov+2025b}, \citet[][triangles]{Boquien+2022}, \citet[][black diamonds]{Fisher+2025} \citet[][black circles]{Burgarella+2025, Burgarella+2026}, \citet[][purple contours]{Nagaraj+2022}, and \citet[dot-dashed yellow line showing the median of their distribution][]{Salim+2018}. \textbf{\thesanzoom galaxies underpredict the attenuation at 1500\,\AA\, even after accounting for unresolved dust in the stellar birth cloud.}}
    \label{fig:attenuation}
\end{figure}

One of the main observable effects of cosmic dust is the reprocessing of UV light into IR. This suppresses the galaxy UV flux and enhances the IR part of the SED, and is often quantified through the attenuation at 1500 \AA, indicated by $A_V$. We compute this value for our simulated galaxies by post-processing them through the SKIRT code \citep{skirt}, as described in detail in a companion paper (Popovic et al., in prep., based on \citealt{Shen+2022}). For the sake of completeness, we briefly summarize our methodology in what follows. 

We produce spectral energy distributions (SEDs) for targets at all snapshots, while for \candidates SEDs are produced only at integer redshifts $z\in[3,14]$. In both cases, we post-process only central galaxies with stellar masses greater than $50$ times the baryonic mass resolution of the corresponding zoom level, yielding a total sample of 7675 galaxies. 
SEDs are produced using the \textsc{SKIRT} code. Each spectrum spans wavelengths $0.05$–$1000\,\mu\mathrm{m}$ sampled with 791 bins, with enhanced resolution around major emission lines. We use $10^7$ photon packets for the stellar (primary) and dust (secondary) emission phases. Stellar radiation is constructed by interpolating between spectral libraries according to stellar particle properties. Stars older than $30\,\mathrm{Myr}$ use the BPASS library \citep{BPASS2018}, while younger stars are modeled with the \texttt{TODDLERS} library \citep{toddlers, toddlers2}, which accounts for \textit{unresolved} \ion{H}{ii} regions, nebular emission, and dust within stellar birth clouds, assuming clouds follow a power-law distribution and that dust  abundance scales linearly with metallicity. \texttt{TODDLERS} then computes analytically the evolution of the gas shell under the impact of newborn stars from collapse to dissolution, including multiple collapse events, photo-ionisation, Lyman-$\alpha$ radiation pressure and external gas pressure. Thus, \texttt{TODDLERS} includes an additional dust component to the simulations. Dust within the interstellar medium is modeled as a mixture of silicate, graphite, and neutral PAH grains, with relative abundances determined by the simulated carbon-to-silicon ratio. Grain size distributions follow the BARE-GR-S model \citep{Zubko+2004} within \textsc{SKIRT}, sampled with 15 bins per species. For each galaxy, SEDs are computed from four tetrahedral viewing angles and combined using an arithmetic mean. We then compute the attenuation as
\begin{equation}
    A_{1500} = -2.5 \log_{10} \left( \frac{F_\mathrm{obs}}{F_\mathrm{int}} \right)\,,
\end{equation}
where $F_\mathrm{int}$ and $F_\mathrm{obs}$ are the intrinsic and observed flux at 1500\,\AA, respectively. 

We plot the simulated $A_{1500}$ values as a function of galaxy stellar mass in Fig.~\ref{fig:attenuation}, using the same color- and symbol-coding as in other plots for the target galaxies at all available redshifts. We show in the background the results for all \candidates galaxies using a grey-shaded histogram. Notice that this is mostly covered by target data, but extends all the way to $\log(\Mstar/\Msun) \sim 9$ with a similar shape. The red foreground histogram and its shaded region represent the median and central 84\% of data in bins of stellar mass. Finally, we show observations from \citet[][black squares]{Markov+2025a,Markov+2025b}, \citet[][triangles]{Boquien+2022}, \citet[][black diamonds]{Fisher+2025}, \citet[][black circles]{Burgarella+2025, Burgarella+2026}, \citet[][purple contours]{Nagaraj+2022}\footnote{We use here the distributions of $A_V$ presented in \citet{Fisher+2025} and compute the $A_{1500}$ using Eq. 5 in \citet{Markov+2025b}. For the parameters $c_1$, $c_2$, $c_3$ and $c_4$ we use their median value in the $z \leq 3$ galaxies in the \citet{Markov+2025a} sample.} and \citet[dot-dashed yellow line showing the median of their distribution][]{Salim+2018}. 

The synthetic SEDs of the \thesanzoom galaxies underpredict the majority of the observed $A_{1500}$ attenuations. Our simulated galaxies lie at the lower end of the data at $M_\mathrm{star} \gtrsim 10^{7.5} \, \Msun$. This is the case despite the contribution of the TODDLERS unresolved birth-cloud dust. When this contribution is removed, the disagreement is much worse. The predicted attenuation without any contribution from TODDLERS is shown in Fig.~\ref{fig:attenuation} using a green dashed histogram. This once again reflects the struggle of reconciling bursty stellar feedback with the build-up of dust, which in our model is easily destroyed or ejected by energetic events. In a forthcoming paper (Popovic et al., in prep.) we will investigate the amount of dust necessary to reconcile our simulations with the observed galaxy properties.

The contribution from the additional \texttt{TODDLERS} dust, while reducing the disagreement with observed attenuations, can impact the other results discussed in this paper. However, we have tested that this is not the case. In fact, the additional dust has a negligible impact on all the scaling relations investigated, since it is always sub-dominant except during the (very-short) ending phase of a starburst, where the simulated dust has been mostly destroyed. However, the fact that it is concentrated around newborn stars has a large impact on the predicted SEDs of galaxies. Thus, we are confident that our result unrelated to the SED modeling are reliable. (We provide more details in Appendix~\ref{app:toddlers}.)

Our results highlight one of the biggest challenges in modeling cosmic dust. The sensitivity of dust properties, including its survival, to the densest and coldest parts of the ISM renders this problem complex. In fact, the cold gas within and around galaxies is sensitive to both the resolution achieved by simulations and the details of stellar feedback. At the same time, this makes the dust properties a powerful probe of physical processes within galaxies. Tools like ALMA (including its proposed upgrade ALMA2040) and NOEMA are key to unveil the formation of the first dust grains and the properties of the first galaxies. 
The results outlined in this Section demonstrate that empirical relations calibrated on observables provide only a partial description of the highly non-linear interplay between gas dynamics, star formation, feedback, and dust processes. Ultimately, modeling dust in simulations not only provides a way to constrain this component of the Universe, but also a promising avenue to better understand the physics of galaxy formation and of the ISM.

\section{Discussion}
\label{sec:discussion}

The results presented in Sec.~\ref{sec:simulated_properties} and Sec.~\ref{sec:observed_properties} depict a clear physical picture. The galaxies simulated in the \thesanzoom project match many of the observed dust properties. However, they consistently struggle to build up enough dust, especially in the most massive galaxies. We have identified the root cause of this issue to lie in the bursty nature of star formation in our galaxy formation model (Sec.~\ref{subsec:dust_mass}). Here, we first draw a comprehensive picture of the link between burstiness and dust survival in our galaxies, and then discuss the implications for our understanding of the formation of the first galaxies.

\subsection{The connection between burstiness and dust survival in the THESAN-ZOOM galaxies}
\label{subsec:burstiness_and_dust_survival}

\begin{figure*}
    \includegraphics[width=\textwidth]{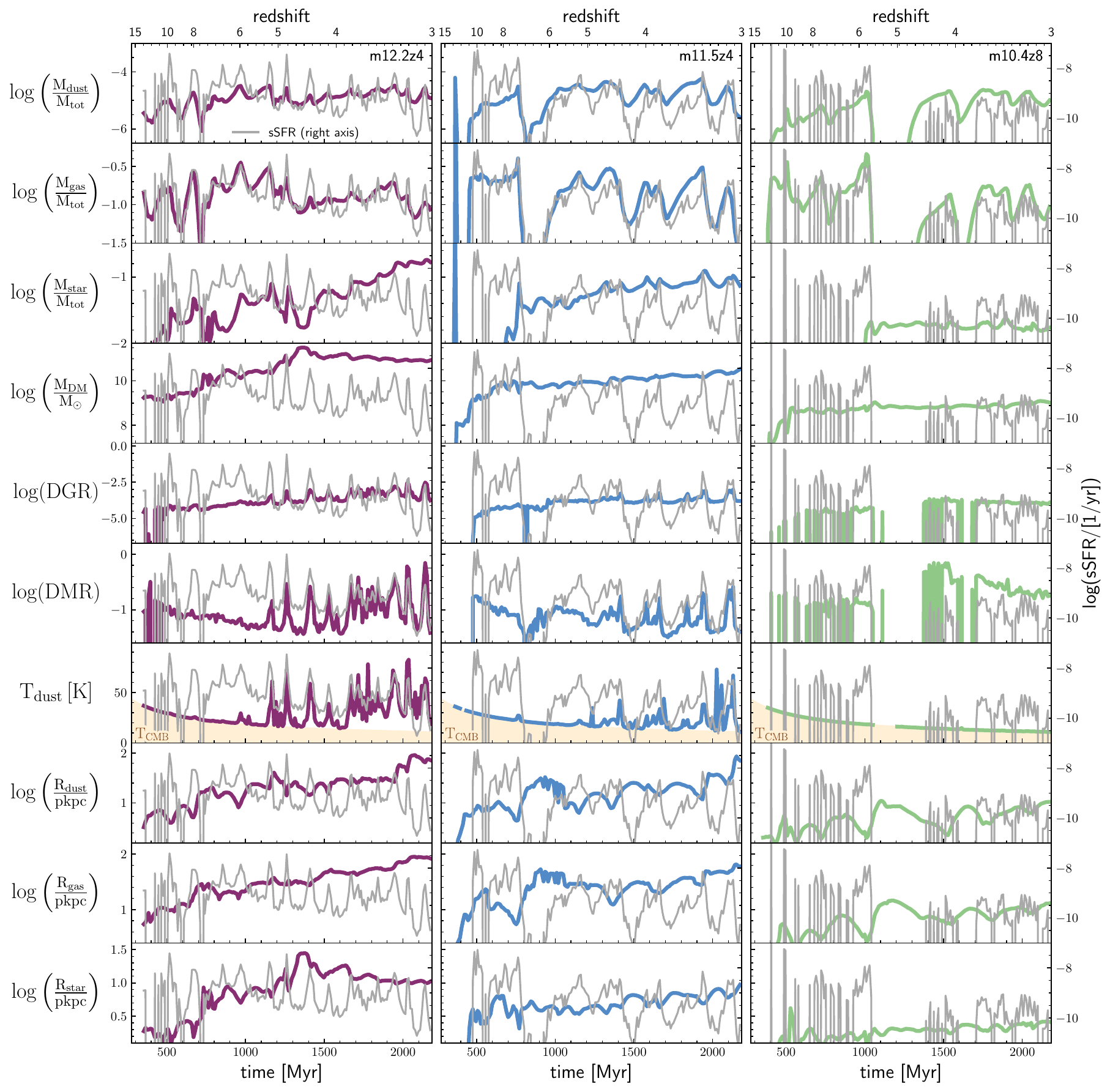}
    \caption{\textbf{Time evolution of physical properties of three sample \thesanzoom galaxies}, namely the \texttt{m12.2z4}, \texttt{m11.5z4} and \texttt{m10.4z8} runs (left to right column, respectively). From top to bottom, we show with coloured thick lines the dust-to-total-mass ratio, gas-to-total-mass ratio, stellar-mass-to-total-mass ratio, DM mass, dust-to-gas ratio, dust-to-metal ratio, IR-luminosity-weighted dust temperature, dust radius, gas radius and stellar radius (solid lines and left-hand-side vertical axis). Radii are computed as twice the half-mass radius of the component. All other quantities are computed using all star-forming cells in the galaxy (see Sec.~\ref{subsec:preliminary}. In each panel, we also show the instantaneous specific star formation rate (sSFR, grey solid line, right-hand-side vertical axis). In the bottom panel, we show with a yellow shading the temperature floor effectively imposed by the CMB. 
    \textbf{Star formation episodes trigger the production of dust, but their feedback rapidly removes gas and dust from the galaxies, often resulting in a net loss of dust regardless of the halo mass.}
    }
    \label{fig:tracks}
\end{figure*}

In Fig.~\ref{fig:tracks}, we show from top to bottom the evolution of the dust-, gas- and stellar-to-total mass ratios, the DM mass $M_\mathrm{DM}$, the dust-to-gas (DGR) and dust-to-metal (DMR) ratios, the IR-luminosity-weighted dust temperature $\Tdust$, as well as the dust, gas, and stellar radii for the target galaxy in the \texttt{m12.2z4}, \texttt{m11.5z4} and \texttt{m10.4z8} runs (left to right column, respectively) using a thick solid line (left y axis). These were chosen as representative examples of the behaviour in different mass bins. In each panel, we also show the instantaneous specific star formation rate (sSFR) of the galaxy (right y axis) with a grey solid line. 
The image clearly shows that the \thesanzoom galaxies cycle through phases of progressive gas growth, followed by a burst of star formation that evacuates a significant amount of gas shortly thereafter, because of supernova explosions of newly-formed massive stars at the end of their lifetimes. This halts star formation and re-starts the cycle. The significance of such mass ejection is highly dependent on the total halo mass, with larger objects (\eg \texttt{m12.2z4}, leftmost column) able to eventually retain most of their gas content once their potential well becomes deep enough, while smaller ones (as \texttt{m10.4z8}, rightmost column) are completely devoid of gas after any star formation event. For a thorough analysis of such behaviour and its implications for observations of high-$z$ galaxies, we refer the reader to \citet{ThZoom_bursty}. For the scope of this paper, we focus on the impact on the dust content of galaxies. 

A careful analysis of Fig.~\ref{fig:tracks} allow us to dissect the exact mechanisms and timeline of the dust cycle in the simulations. The first two rows of Fig.~\ref{fig:tracks} show that gas and dust tend to increase together prior to a starburst event and, similarly, to be removed after its peak\footnote{While our model treats dust as a property of gas cells and therefore does not allow for different dynamics for these two components, we expect this to play only a minor role in the removal of dust from the galaxy. In fact, not only is the momentum injected by the supernova expected to couple to dust as well as to gas, but SN explosions are known for efficiently destroying dust through sputtering. Finally, our model also accounts for dust production through condensation in supernova ejecta, adding a \textit{source} of dust during supernova explosions that can, in principle, counterbalance the depletion caused by the SN explosion itself.}. 

Turning our attention to the evolution of the dust-to-gas and dust-to-metal ratios (fifth and sixth rows from the top) reveals that during the initial phases of accretion, the dust and gas grow simultaneously (\ie the DGR remains constant) due to the influx of dust-rich material either from the circumgalactic medium or through mergers. Eventually, the dust mass growth outpaces that of the gas (the DGR increases). This occurs once the gas density is large enough to render the condensation of gas-phase metals into dust efficient (due to the high densities reached), as can be seen by looking at the DMR, which increases. The same gas growth, however, soon triggers the formation of new stars (third row in the Figure). Their feedback quickly heats up the gas, disrupting further star formation and removing dust (either through destruction or ejection). Interestingly, during the final phases of these starbursts, both the DGR and DTM ratios revert to values close to those pre-starburst, indicating that dust-rich gas is preferentially affected by dust removal processes. This is a consequence of the fact that such gas is typically co-spatial with the newborn stars, as dust is formed through ISM accretion in the same dense gas that produces stars. Therefore, stellar feedback preferentially and efficiently destroys such gas reservoirs. Incidentally, this is also the underlying explanation for the evolution of dust temperature in our simulated galaxies (fourth row from the bottom). The proximity to newborn stars entails that their strong UV radiation can efficiently heat up the dust, which quickly dominates the  IR luminosity of a galaxy (and, therefore, the IR-luminosity-weighted temperature shown here), but also that such hot dust is preferentially removed. The dust surviving the starburst-induced stellar feedback is thus preferentially far from star formation sites and therefore colder. 

The rapid dispersal of gas around newborn stars that we observe in our simulations is in overall agreement with observations of stellar clusters, finding that these objects remain embedded in their birth cloud for $3$--$5$ Myr \citep[\eg][]{Leisawitz+1989,Whitmore&Zhang2002,Lada&Lada2003,PortegiesZwart+2010,Kim+2021,Schinnerer&Leroy2024}. Recently, simulations have started to probe this process \citep{Wainer+2026}, revealing that this process is driven by the most massive stars.

\subsection{Implications for the UV luminosity function and escape fractions}
\label{subsec:implications_uvlf}

Reconciling bursty feedback and dust survival is not trivial. Simply increasing the dust formation rate, or decreasing the dust destruction rates does not affect this result much, since stellar feedback during and after starbursts evacuates virtually all the gas present in the galaxy (see Fig.~\ref{fig:tracks} and related discussion). Another possibility is to assume that dust is formed in unresolved regions, as the stellar birth cloud. If these reach sufficiently-large densities, they might survive the strong stellar feedback and preserve part of the dust content during starbursts. It is important to note that the choice of star formation efficiency affects the amount of cold gas able to survive, as more efficient star formation removes more effectively cold gas clumps as they are turned into stellar particles. As an initial step toward testing the impact of this parameter, we have analyses three runs with variable star formation efficiency (linearly increasing from 1\% at gas density $n_\mathrm{H} = 10$ cm$^{-3}$ to 100\% at $n_\mathrm{H} \geq 10^3$ cm$^{-3}$). This model variation is available only for the \texttt{11.1z4}, \texttt{10.4z4}, \texttt{10.4z8}, \texttt{9.7z4} and \texttt{9.7z8}. We do not find significant differences in the dust survival. However, the high non-linearity of the dust evolution demands a dedicated study to establish whether bursty feedback can be reconciled with dust survival. 

Intriguingly, bursty star formation has been widely invoked as a potential explanation for the large number of UV-bright galaxies observed by JWST \citep[\eg][]{Bouwens2023,Harikane+2023, Leung+2023, Perez-Gonzalez+2023b, Donnan+2023, Donnan+2024, McLeod+2024, Chemerynska+2024, Robertson+2024, Whitler+2025,Semenov+2025}. UV variability can temporarily scatter moderate-mass galaxies into the bright tail of the luminosity function \citep[although the magnitude of the expected UV scatter remains model-dependent and sensitive to feedback physics and burst duty cycles;][]{Shen+2023, Sun+2023, Kravtsov+2024} and therefore accommodate a larger number of UV-bright objects within the standard $\Lambda$CDM paradigm. 
In fact, a small but increasing set of galaxies with suppressed star formation at early times (``mini-quenched'') has been identified \citep[\eg][]{Looser+2024, Baker+2025}. 

It naturally follows from the aforementioned considerations that our results place constraints on the viability of bursty star formation as an explanation for the enhanced UV luminosity function (UVLF). Our simulations suggest that a bursty galaxy formation process is unable to retain sufficient amounts of dust to match observations. However, observations of the dust content of galaxies are extremely limited at the redshifts where measurements of the UV luminosity function exceed model predictions. 
For this reason, pushing observations of the dust properties to fainter and/or higher-redshift objects is mandatory to advance our understanding of the assembly of the first galaxies. Programs like PHOENIX\footnote{PHOENIX is a recently-selected ALMA Large Program that will observe dust continuum emission in 15 intrinsically luminous galaxies at $8<z<15$ that have ancillary JWST data, including their [OIII]88$\mu$m emission.} will be instrumental to this end. 
Should similar programs find large dust reservoirs in $z \gtrsim 10$ UV-bright galaxies, it would indicate that bursty star formation is unlikely to be the explanation for the over-abundance of UV-bright galaxies in the first few hundred million years of the Universe, unless it can be achieved while limiting at the same time the impact of stellar feedback on the dust-rich gas. 

At lower redshifts ($z \lesssim 8$), our simulations over-predict the UV luminosity function at $\mathrm{M}_\mathrm{UV} \lesssim -18$, indicating a lack of dust attenuation (see also Sec.~\ref{subsec:attenuation}). At these redshifts, however, bursty star formation is not necessary to explain the observed abundance of UV-bright galaxies. Thus, if the burstiness of star formation is the underlying cause of the elevated UVLF observed, then it must quickly drop at $8 \lesssim z \lesssim 10$.

As always with numerical works, our conclusions are model-dependent. However, the prediction that bursty star formation removes a large fraction of dust from the galaxy is remarkably independent of the details of our dust model, because, regardless of how much or how quickly dust is formed, it will be removed during a starburst event. Thus, the incompatibility between bursty star formation and dust content predicted by our model is resilient to the (highly uncertain) dust physics. 
We have also tested the physics variations of the \thesanzoom model (which affect the galaxy formation model but not dust), as well as a new variation of the dust module (see Appendix~\ref{app:newdust}), and found that, while the burstiness of galaxies and the dust content \textit{between} starburst episodes are altered, the lack of dust in the post-starburst phase remains. 

Finally, our results have implications also for the ionizing escape fraction in high-redshift galaxies. The burtiness of star formation has been connected to the escape of Lyman continuum (LyC) radiation by a number of works, both theoretical \citep[\eg][]{Ma+2018, Smith+19, Kimm+2021, Katz+2023, Ceverino+2023} and observational \citep[\eg][]{Izotov+2016_nat, Izotov+2016_mnras}. Therefore, we predict that LyC leakers are preferentially in a dust-poor phase.

\subsection{Comparison with other works}
\label{subsec:comaprison_models}

A number of models have attempted to understand the properties of cosmic dust and its role in galaxy evolution and in explaining observations. We have discussed many of them in the relevant sections throughout the paper. In the following we focus on two recent works that made predictions testable or complemented by our simulations. 

\subsubsection{Attenuation-free model \citep{Ferrara+2023}}
\citet{Ferrara+2023} proposed what became known as the attenuation-free model (AFM) to explain the unexpected overabundance of UV-bright massive galaxies at redshift $z>10$ (dubbed `blue monsters' by these authors). In this model, the large sSFR of these galaxies triggers powerful radiation-driven outflows, dispersing dust out to kpc scales and, therefore, making them transparent to UV radiation. This also explains their blue spectral slopes ($\beta \lesssim2$) and negligible UV attenuation, ALMA non-detections of their dust continuum, enhanced Lyman-$\alpha$ visibility \citep[through outflow-induced velocity offsets facilitating its survival through the neutral IGM][]{Ferrara+2024} and anti-correlation between Eddington ratio and dust UV optical depth \citep{Fiore+2023}. 

Our simulations achieve the same end result, but through stellar feedback instead of just radiation pressure on dust. Notably, this releases a key assumption of the AFM, namely that radiation pressure on dust is boosted by a factor $100$ or more because of multiple scatterings of IR photons. This effect is not explicitly included in our simulations, although our model produces an increased IR radiation pressure at large optical depths, \citep[see Section 3.2.2 of][]{ArepoRT}. 
A key difference in the outcome is, however, that our bursty stellar feedback destroys a significant fraction of the dust content of our galaxies, while the rest is mostly expelled (often to $\gtrsim3$ times the galaxy effective radii). \citet{Ferrara+2025_bluemonster} use an analytical model for dust formation and destruction to infer the dust-to-stellar ratio in observed `blue monsters' under the assumption of no outflows. They find values approximately $100$ times larger than in observed galaxies, and thus conclude that dust ejection must dominate over destruction. However, we find values of the dust-to-stellar ratio that are comparable, although slightly larger, to those found in the `blue monsters' (Fig.~\ref{fig:Mdust_vs_Mstar} and Fig.~\ref{fig:dust-to-star_vs_SFR}).

There are a number of possible reasons why our simulations and the AFM model disagree on the fate of dust. First and foremost, our simulation suite contains only a small number of galaxies comparable (in terms of redshift, stellar and dust masses, and SFR) to the `blue monsters' discussed by \citet{Ferrara+2023}. Second, our simulation suite might feature a lower IR radiation pressure on dust, either due to lack of resolution or because of optimistic assumptions in the AFM, therefore requiring stellar feedback to drive outflows. A third possibility is that the analytical model used by \citet{Ferrara+2025_bluemonster} to determine the dust-to-stellar ratio might not hold in a realistic scenario. Two assumptions are particularly different from the results of our simulations, namely: the assumption that galaxies accrete gas in a continuous and smooth fashion (while our and many other simulation indicate that merger events dominate the mass accretion history at these early times) and the assumption that there is no outflow from the galaxy. The first one might be problematic because such `monsters' are likely in very biased regions of the cosmic web and, therefore, are likely to undergo an accelerated formation process and experience an unusually-large number of mergers. The second approximation instead could artificially boost the gas reservoir available and, thus, dust production. This, in turn, could artificially push the required dust destruction coefficient into unphysically-large values. In our simulations, even when dust is preferentially destroyed during a starburst, gas is nevertheless expelled. This is the main physical mechanism that actually stops the starburst. 

Overall, our results support the general picture of the AFM, but the ultimate fate of dust requires further investigation. It should also be noted that, as discussed in Sec.~\ref{sec:simulated_properties}, our galaxies seem to under-produce dust in massive systems with respect to observations at $z \lesssim 6$. Should dust survive the starburst phase, as predicted by the AFM, and be re-accreted by the galaxies, this discrepancy could be reduced or even resolved. This process, however, seems at odds with a bursty star formation history (see Sec.~\ref{subsec:burstiness_and_dust_survival}).

\subsubsection{The first dust grains \citep{Burgarella+2025, Burgarella+2026}}
Recently, \citet{Burgarella+2025, Burgarella+2026} suggested the existence of two distinct populations of galaxies, based on observations that are part of the CEERS program. Specifically, these authors suggest the existence of a dust-rich and a dust-poor population, with the latter being formed by galaxies undergoing their first dust-enrichment episode. While these authors interpret their observations as indicative of the emergence of the first dust grains, we instead provide an alternative explanation. Should this dichotomy in dust properties be confirmed, we predict that the dust-rich population has either: \textit{(i)} systematically quieter star formation histories than the dust-poor one, or \textit{(ii)} traces a pre-starburst phase while the dust-poor population traces post-starburst galaxies The existence of distinct populations could arise from short timescales of transition between different states or diverging evolution due to physical triggers (\eg AGNs, critical halo mass, etc.).

It should be noted, however, that \citet{Burgarella+2025} infer $M_\mathrm{gas} / (M_\mathrm{gas} + M_\mathrm{stars}) > 90$\% for all galaxies in their sample. In \thesanzoom galaxies, such large gas fractions are achieved shortly before a starburst event, but are not sustained throughout the galaxy life (see the discussion in Sec.~\ref{subsec:burstiness_and_dust_survival}). Thus, if confirmed, these measurements would indicate drastically different physical conditions in observed and simulated galaxies.

\subsubsection{Bursty star formation from turbulence-driven star formation \citep{Semenov+2025}}
\citet{Semenov+2025,Semenov+2025b} developed and investigated a model of turbulence-driven star formation. This is able to achieve a bursty star formation history in galaxies without relying on stellar feedback ejecting the ISM gas and halting star formation. It is therefore possible that such model could match the observed UVLF while accommodating a milder stellar feedback. This, in turn, could aid the survival of significant amounts of dust in the galaxies. However, this possibility relies on the detailed balance between clustered (in time and space) stellar feedback and dust evolution, thus rendering dedicated numerical studies necessary to assess whether such model could explain the observed galaxy properties.

\subsection{Model limitations}
\label{subsec:limitations}
Our simulations include a remarkable range of physical processes. Nevertheless, limitation in modeling or resolution remain. As discussed above, it is possible that our finite resolution prevents us from capturing the densest gas clumps. This could impact the topology of star formation, and therefore its impact on the galaxy. Moreover, since dust growth is more efficient at higher gas densities, these very dense clumps might produce a faster recovery of dust mass following a starburst phase. In fact, some simulations \citep[\eg][]{Aoyama+2018, colibre_dust} explicitly assume that all `cold' gas ($n_\mathrm{gas} \sim 0.1$ cm$^{-3}$, $T_\mathrm{gas}\lesssim10^4$ K) is in fact in an unresolved denser and colder phase ($n_\mathrm{gas} \gtrsim 10$ cm$^{-3}$, $T_\mathrm{gas} \lesssim 100$ K), which boosts the dust growth. 
However, we expect this dust growth channel to be of minor importance in our model, since stellar feedback during starburst removes gas from the galaxy, including the potentially unresolved clumps. 
On the other hand, it is also conceivable that these dense clumps could partially shield the ISM from the intense stellar feedback responsible for the depletion of dust after starburst, thus mitigating its disruptive effects and enabling gas and dust survival. 

Assumptions concerning dust modeling are particularly relevant for the results discussed in this Paper. In particular, we model dust using a single effective grain size. While numerically efficient, this might prevent us from capturing physical processes related to dust size evolution. Specifically, the size of dust grains affects how efficiently they accrete metals and how they affect the UV light. It can also affect how they react to feedback, with larger grains requiring more energy to be accelerated (and thus ejected from the galaxy), but also conserving their momentum for longer \citep{Hu+2019,Slavin+2020,Priestley+2022}. 

Multiple studies have modelled the grain size distribution, although often without the resolution or breadth of physical processes featured in this Paper. As an example of a recent work of this kind, \citet{Narayanan+2025,Narayanan+2026} found that in their simulations dust obscuration is driven by high grain-grain shattering rates. Moreover, their `blue monsters` galaxy are not completely dust depleted, but have a top-heavy grain size distribution that renders them optically thin to the UV light.

\section{Summary and Conclusions}
\label{sec:conclusions}
We have presented a systematic analysis of cosmic dust in the \thesanzoom project, a suite of radiation-hydrodynamical simulations producing a multi-phase ISM and bursty star-formation. The simulations include an on-the-fly model for dust formation, growth and destruction, as well as coupling between dust physics and radiative transfer, thus enabling self-consistent predictions of dust temperatures. 

Our main conclusions are as follows.

\begin{itemize}
    \item[1.] \thesanzoom galaxies fit well in multiple observed scaling relations between dust content, metallicity and stellar mass. In particular the dust-to-gas and  dust-to-metal ratios as functions of gas metallicity show an excellent agreement with most datasets. The simulated galaxies contain less dust mass at a given stellar mass when compared to most observations, although the latter are biased towards IR-bright (and thus dust-rich) galaxies. Intriguingly, our simulations seem to reproduce the `blue monsters' in \citet{Ferrara+2025_bluemonster} and the dust-poor population of \citet{Burgarella+2025}.
    
    \item[2.] The dust-to-stellar ratio is in agreement with observations for low specific star formation rates, but flattens below the observed values at higher sSFRs. We interpret this as an indication that dust is removed during starbursts.

    \item[3.] The predicted dust temperatures are in broad agreement with observed values. Our simulated dust is heated above the CMB temperature only for short amounts of time during starbursts. These become increasingly stronger with time (until $z=3$ where our simulations stop), driving up the average dust temperature in the sample. Our simulations predict a correlation between the dust temperature and the distance from the galaxy main sequence, that we tentatively find in observed objects. 

    \item[4.] The extent of the dust distributions within individual objects changes dramatically with time, with excursions up to an order of magnitude within a few tens of Myr. On average, dust is more extended than stars and less than the gas. 

    \item[5.] The \thesanzoom galaxies show a broad range of UV--IR morphologies. We find widespread offsets between the peaks of the two distributions, even after accounting for observational effects. The distribution of UV--IR offsets is in good agreement with those observed by the ALPINE and REBELS programs. Our results demonstrate that the commonly-used scaling relations can hide the highly-non-linear interplay with ISM physics. Thus, dust provides an important avenue to constrain ISM properties.

    \item[6.] The simulated galaxies show a severe lack of attenuation at $1500$ \AA{} compared to observed ones. This is due to stellar feedback destroying dust in the star-formation sites, thus allowing UV radiation to escape. Accounting for unresolved stellar-birth-cloud dust (using the TODDLERS library) brings the simulated attenuation closer to the observed data without significantly changing the scaling relations investigated because of the small amount of dust injected.
    
    \item[7.] We predict that dust-poor galaxies have high escape fractions of ionising photons, since in our simulations this evolutionary phase is driven by the ejection of gas (and destruction of dust) by the intense stellar feedback following a starburst.

    \item[8.] Our results are similar to the attenuation-free model of \citet{Ferrara+2023} for what concerns UV-bright $z\sim10$ `blue monsters', but with different physical mechanism at play. In our simulations, these objects are created by stellar feedback removing dust from star-forming regions. Unlike the original attenuation-free model, radiation pressure does not play a relevant role. This has implications for the fate of dust, that in our model is mainly destroyed by such feedback events and therefore can not be re-accreted by the galaxies. Despite this difference, we find that the dust-to-stellar ratio in our simulations is in agreement with observed values (although slightly higher). These results, however, are accompanied by a lack of dust at lower redshifts; while we expect that a reduced burstiness of star formation would allow our simulations to match observed  data at $z \lesssim 7$, further studies are needed to confirm this.

    \item[9.] Dust is a powerful diagnostic of the star-formation and feedback cycle. Extending dust observations to fainter and higher-redshift systems, and improving constraints on small-scale dust survival and re-accretion, will be essential to discriminate between competing models for the first generations of galaxies.
    
\end{itemize}

The main take-away from this paper is that, within our galaxy formation and dust models, bursty star formation is incompatible with a substantial dust reservoir surviving over prolonged periods of time. 
During starbursts, feedback-driven outflows remove gas and dust efficiently; after the burst, galaxies enter dust-poor phases where the low density of gas prevents efficient dust growth. This phase lasts until the beginning of the next starburst cycle, where gas inflow increases densities and dust content, but also triggers star formation that quickly removes gas and dust. Although our study has limitations, this physical mechanism is likely common to multiple galaxy formation models. Thus, we it should be thoroughly explored in a wide variety of numerical models in order to conclusively determine the (in)compatibility of bursty star formation and dust survival.

\section*{Acknowledgements}
We are thankful to Andrea Ferrara, Denis Burgarella, Anand Utsav Kapoor, Vladan Markov, Joe Lewis, Pierre Ocvirk, Meghana Killi and Michele Ginolfi for fruitful discussions as well as for sharing their data with us. 

EG is supported by the JSPS KAKENHI grant ILR 23K20035. EG acknowledges support from the CANON Foundation Europe during part of the work resulting in this publication. NY acknowledges the support of the Japanese Society for the Promotion of Science through their Specially Promoted Research 24H00004. 
KN acknowledges support from JSPS KAKENHI grant 20H00180, 24H00002, 24H00241, JP25K01032, and the JSPS International Leading Research (ILR) project, JP22K21349. 
KN also acknowledges support from the Kavli IPMU, the World Premier Research Center Initiative (WPI), UTIAS, the University of Tokyo. 
WM thanks the Science and Technology Facilities Council (STFC) Center for Doctoral Training (CDT) in Data Intensive Science at the University of Cambridge (STFC grant number 2742968) for a PhD studentship. 
XS is supported by the NASA theory grant JWST-AR-04814. LK acknowledges the support of a Royal Society University Research Fellowship (grant number URF$\backslash$R1$\backslash$251793). 
AS acknowledges support through JWST AR-08709.

We gratefully acknowledge the Gauss Centre for Supercomputing e.V. (\url{www.gauss-centre.eu}) for funding this project by providing computing time on the GCS Supercomputer SuperMUC-NG at Leibniz Supercomputing Centre (\url{www.lrz.de}), under project pn29we. 
We acknowledge the use of the XD2000 system at the Center for Computational Astrophysics of the National Astronomical Observatory of Japan, and of computational resources from the Kavli IPMU and the MIT Office of Research Computing and Data that have contributed to the research results reported within this paper.

We are thankful to the community developing and maintaining software packages extensively used in our work, namely: \texttt{matplotlib} \citep{matplotlib}, \texttt{numpy} \citep{numpy}, \texttt{scipy} \citep{scipy}, \texttt{cmasher} \citep{cmasher}, \textsc{AREPO} \citep{Arepo, Arepo-public}. 

\section*{Author contributions}
We list here the authors contributions following the CRediT\footnote{\url{https://www.elsevier.com/researcher/author/policies-and-guidelines/credit-author-statement}} system. EG: conceptualization, methodology, software, formal analysis, validation, writing -- original draft, writing -- review and editing, visualization, supervision, project administration. FP, RK, AS, EP: software, review and editing. LH, LK, KN, NY, CP, WM, XS, ST, MV: review and editing.

\bibliographystyle{mnras}
\bibliography{bibliography}

\appendix

\section{Numerical convergence}
\label{app:numerical_convergence}

\begin{figure}
    \includegraphics[width=\columnwidth]{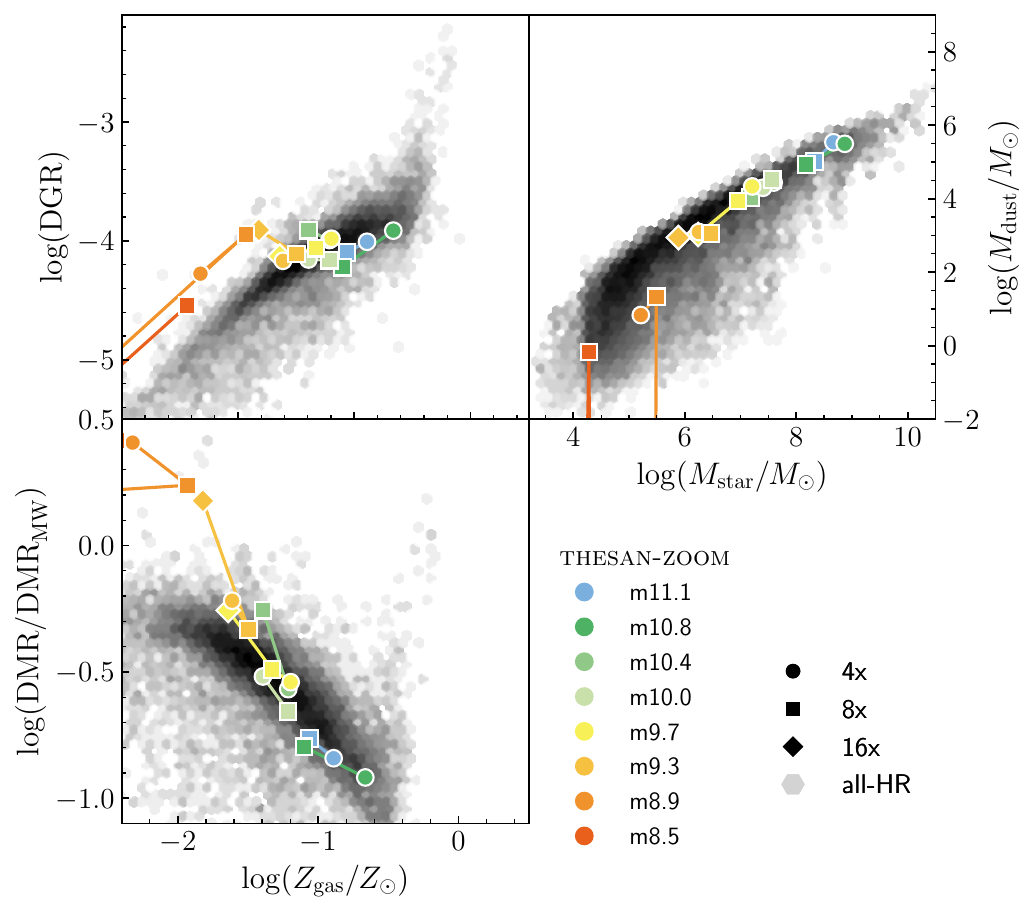}
    \caption{\textbf{Numerical convergence test} for the dust properties of \thesanzoom galaxies. Each panel shows a different scaling relation, namely: dust-to-gas ratio (top left) and dust-to-metal ratio (bottom left) as functions of gas metallicity, and dust mass as a function of stellar mass (top right). Symbols joined by lines show values computed for the same target galaxy at different resolution levels (circles for \zf, squares for \ze and diamonds for \zs). To mitigate the impact of stochasticity on our results, we show the median values computed over a redshift window $3 \leq z < 4$. The background shaded region shows the distribution of all other (targets and \candidates) galaxies at $z=3$. \textbf{\thesanzoom galaxies show limited variations with resolution, especially at the high-mass end accessible to observations. } }
    \label{fig:numconv}
\end{figure}

In this Appendix we test the degree of numerical convergence of our simulations. To this end, we revisit some of the scaling relations investigated in Sec.~\ref{sec:simulated_properties}. In Fig.~\ref{fig:numconv} we show the dust-to-gas ratio (top left) and dust-to-metal ratio (bottom left) as functions of gas metallicity and dust mass as a function of stellar mass (top right). Each colored symbol represents one of the simulated galaxies where multiple resolution levels are available, with circles indicating the base zoom level (\zf), while squares and diamonds represent an improvement by $8$ (\ze) and $64$ (\zs) times in mass resolution, respectively. To mitigate the impact of stochasticity (e.g. slightly different times of star formation, merger, etc. in different zoom levels), we report the median values computed over the redshift window $3 \leq z < 4$. For reference, the shaded region  in the background of each panel shows the distribution of all other (targets and \candidates) simulated galaxies at $z=3$. Notice that we use significantly smaller vertical ranges in the top left panel for visual clarity. However, this exaggerates the apparent difference between different resolution levels. 

The convergence of dust properties in \thesanzoom galaxies is good overall (rarely are two same-color points found very far from each other). The largest differences are found in the smallest/most metal-poor galaxies, which are more susceptible to the details of individual star formation and feedback events. 
All points roughly follow the distribution of other (targets and \candidates) simulated galaxies, which we plot as a shaded background with opacity scaling with the number density. There is no evidence that numerical resolution affects such relations. In fact, simulated galaxies move \textit{along} the relation outlined by the background distributions rather than across/away from it. 
Overall, we believe that our galaxy sample has a satisfactory level of numerical convergence that allows us to trust the results presented in this and other papers.

\section{Gas temperature in star-forming cells}
\label{app:newdust}

\begin{figure}
    \includegraphics[width=\columnwidth]{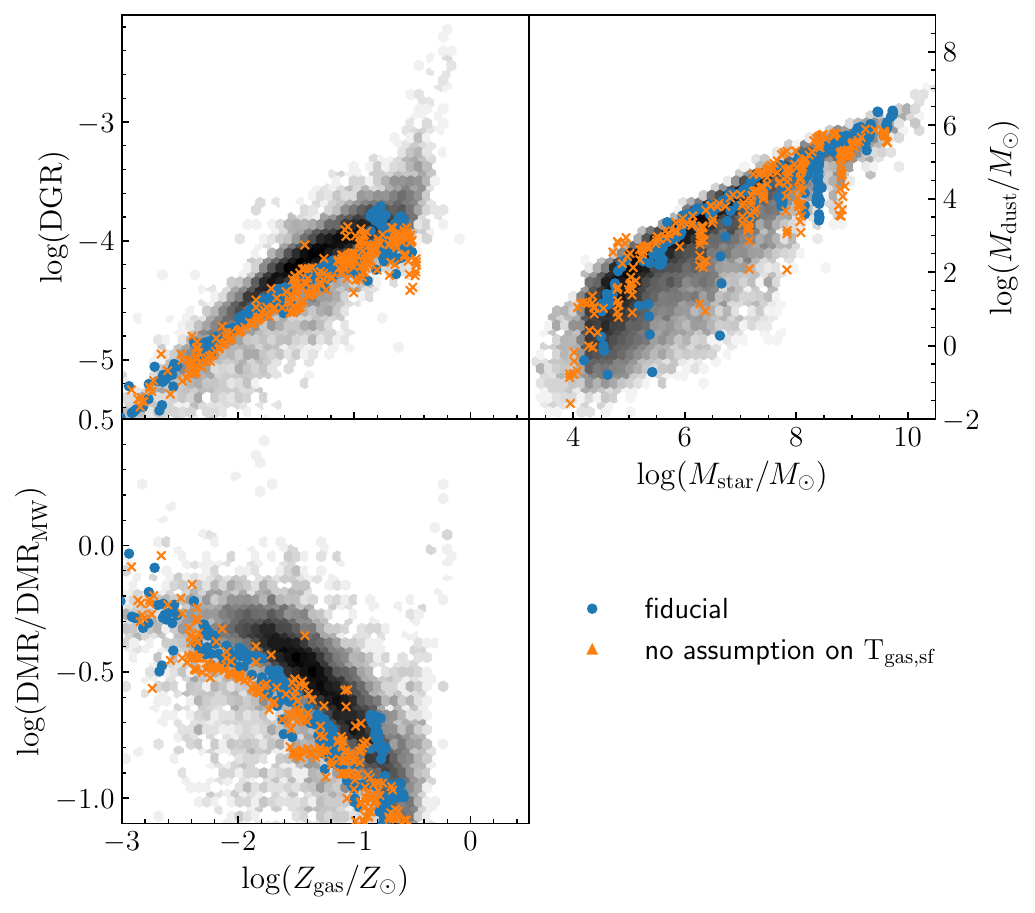}
    \caption{\textbf{Impact of our assumption on the gas temperature} in star-forming regions. Each panel shows a different scaling relation, namely: dust-to-gas ratio (top left) and dust-to-metal ratio (bottom left) as functions of gas metallicity, and dust mass as a function of stellar mass (top right). Blue circles show the fiducial model, while orange crosses show the results directly employing the gas temperature in the dust accretion timescale. We show results for the following targets: \texttt{m12.2z4}, \texttt{m11.5z4} and \texttt{m10.0z8} The background shaded region shows the distribution of all other (targets and \candidates) galaxies at $z=3$. \textbf{Assuming a fixed gas temperature in the dust accretion timescale, as in the original \thesan model, does not impact the dust properties of the simulated galaxies.} }
    \label{fig:newdust}
\end{figure}

In this Appendix we show that assuming a fixed gas temperature for the dust accretion timescale within star-forming cells does not affect the dust properties of our simulations. We have re-run the \texttt{m12.2z4}, \texttt{m11.5z4} and \texttt{m10.0z8} haloes removing the aforementioned assumption and instead directly use the cell gas temperature in the computation of the accretion timescale. 
In Fig.~\ref{fig:newdust} we show the dust-to-gas ratio (top left) and dust-to-metal ratio (bottom left) as functions of gas metallicity and dust mass as a function of stellar mass (top right) for these haloes, using blue circles for the fiducial dust model and orange crosses for the modified one. The distributions of all targets and \candidates in the \thesanzoom suite are shown by shading in the background. The image shows clearly that the assumption of fixed dust temperature does not impact the dust properties, since the distribution of blue points and orange crosses overlap.

\section{Impact of TODDLERS dust on scaling relations}
\label{app:toddlers}

Here, we demonstrate that the inclusion of birth-cloud dust from the \texttt{TODDLERS} library does not significantly modify the scaling relations investigated in Sec.~\ref{sec:simulated_properties}. We compute the amount of dust injected as (Kapoor, private communication):
\begin{equation}
    M_\mathrm{dust, TODDLERS} \approx 3.6 \times 10^6 \times \frac{Z}{0.02} \times \mathrm{SFR}~.
\end{equation}
This formula assumes that a starburst is exactly $30$ Myr in duration, a dust-to-metal ratio of 45\% and a stellar-to-gas mass ratio of $\eta = 0.025$, and should be exact within 20\%. We therefore use this equation to estimate the additional dust mass to inject in each gas cell of the simulated galaxy. At the same time, we decrease the gas-phase metal mass in the cell by an equal amount. This is justified by the assumption that the unresolved dust is purely due to small scale dense structures not resolved at our resolution scale. Therefore:
\begin{eqnarray}
  M_\mathrm{dust, effective}   =& M_\mathrm{dust, simulation}   &+ M_\mathrm{dust, TODDLERS} \\
  M_\mathrm{metals, effective} =& M_\mathrm{metals, simulation} &- M_\mathrm{dust, TODDLERS}
\end{eqnarray}

\begin{figure}
    \includegraphics[width=\columnwidth]{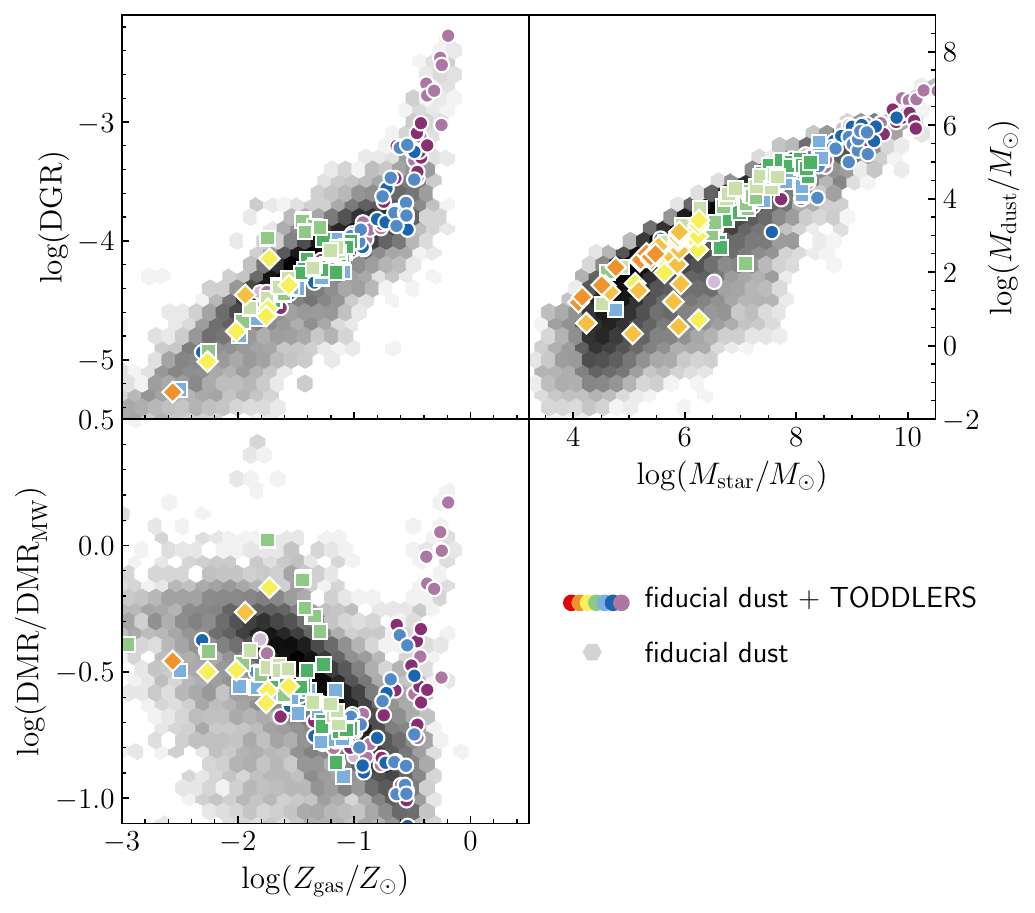}
    \caption{\textbf{Impact of the birth-cloud unresolved dust included by the TODDLERS library} on different scaling relations, namely: dust-to-gas ratio (top left) and dust-to-metal ratio (bottom left) as functions of gas metallicity, and dust mass as a function of stellar mass (top right). Colored circles show the target dust content when TODDLERS dust is included, while the background shaded region shows the distribution of all (targets and \candidates) galaxies at $z=3$. \textbf{The inclusion of unresolved dust in the stellar birth cloud does not significantly change the dust properties of the simulated galaxies.} }
    \label{fig:toddlers}
\end{figure}

We show in Fig.~\ref{fig:toddlers} that the inclusion of additional birth-cloud dust from the TODDLERS library does not significantly affect the scaling relations investigated in the main text. Specifically, we show the dust-to-gas ratio (top left) and dust-to-metal ratio (bottom left) as functions of gas metallicity and dust mass as a function of stellar mass (top right) for all target haloes, where the dust properties have been recomputed to include the contribution of dust injected by the TODDLERS library, as described above. The distributions of all targets and \candidates in the fiducial dust model are shown by a background shading. The image clearly shows that the additional unresolved dust does not significantly affect the scaling relations. 

\end{document}